\documentclass[letterpaper,10pt]{book}%
\usepackage[Conny]{manual}
\usepackage{epsfig}
\usepackage{symb} 
\usepackage{epsf}
\usepackage{amsmath}
\usepackage{amssymb}
\usepackage{graphicx}
\usepackage{gensymb}
\usepackage{defs}
\usepackage{vmargin}
\usepackage{textcomp}
\usepackage{color}
\usepackage{mathrsfs}
\setmarginsrb{30mm}{20mm}{20mm}{25mm}{0pt}{10mm}{0pt}{10mm}

\def\be{\begin{equation}}
\def\ee{\end{equation}}
\def\bea{\begin{eqnarray}}
\def\eea{\end{eqnarray}}

\def\om{\Omega_m}
\def\ode{\Omega_{DE}}
\def\ok{\Omega_k}
\def\p{\partial}
\def\({\left(}
\def\){\right)}
\def\intzinf{\int_z^\infty}
\def\integ0a{\int_0^a}

\def\versionnum{Version 2.0}
\def\name{Fisher4Cast}
\def\fig_dir#1{#1}

\def\da_0/do_k{\(-\frac{1}{2} \frac{a_0}{\ok}\)}
\def\de/dE/do_m{\frac{1}{2} \frac{1}{E} \left [ \left ( \frac{a}{a_0} \right ) ^{-3} - f\right ]}

\def\df/da_0{\left( \( \frac{3(1+w_0+w_a)}{a}  \(\frac{a_0}{a}\)^{3(1+w_0+w_a)-1} \times \exp \left \{3w_a\left(\frac{a_0}{a}-1 \right) \right \}\)  +  \(\frac{a_0}{a}\)^{3(1+w_0+w_a)} \times 3\frac{w_a}{a} \exp \left \{3w_a\left (\frac{a_0}{a}-1 \right ) \right \}\right)}

\def\df/do_k{\(\df/da_0\) \(\da_0/\do_k\)}

\def\dfdw0#1{f(#1) 3 \ln(1+#1)}

\begin{document}

\title{\name~\versionnum\vspace{0.3 in}\\Users' Manual}

\author{Bruce A. Bassett$^{1,2}$, Yabebal Fantaye$^{1,2,3}$, Ren\'{e}e Hlozek$^{1,2,4}$, Jacques Kotze$^{2}$\\
\\
$^1$ {\em South African Astronomical Observatory} \\ {\em Observatory, Cape Town, South Africa} \\
 \\
$^2${\em Department of Mathematics and Applied Mathematics, University of Cape Town}\\{\em Rondebosch, 7700, Cape Town, South Africa }\\
 \\
$^3$ {\em Astrophysics Sector, SISSA}\\{\em Via Beirut 4, 34014 Trieste, Italy}\\
\\
$^4$ {\em Department of Astrophysics, Oxford University}\\{\em Denys Wilkinson Building, Keble Road, OX1 3RH, United Kingdom}
}
\date{}
\maketitle
\let\cleardoublepage\clearpage
\pagestyle{empty}
\tableofcontents
\renewcommand\pagestyle{plain}
\setcounter{page}{1} \pagenumbering{arabic}
\chapter{The Code}
\begin{figure}[htbp!]
\begin{center}
\includegraphics[width=0.4\textwidth]{\fig_dir{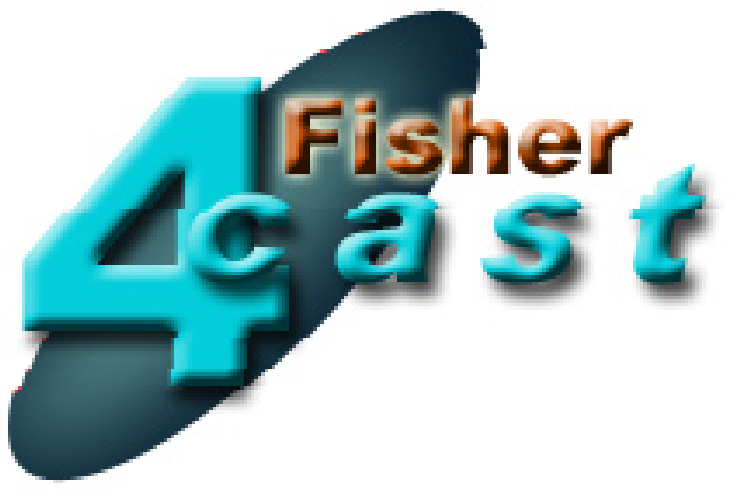}}
\end{center}
\end{figure}
Here we introduce \name{}, a code developed for general Fisher Matrix analysis. In addition to this manual there is also a release paper for \name~which provides more background detail and cosmological applications as well as sample code for generating the plots in the release paper \cite{f4c}. This Users' Manual explains how to install and run the code, both in the Command-line and the Graphical User Interface (GUI), which is coded for cosmology.  The structure of the code is described in detail, as well as the various tests performed during the development of the code.
\section{Introduction}
The Fisher Matrix translates errors on observable quantities measured in a survey into constraints on parameters of interest in the underlying model. As such, it is the elegant way of doing propagation of errors to the case of multiple measurements and many parameters \cite{tegmark}.

In contemporary cosmology, Fisher matrices are used to forecast parameter constraints from a proposed survey, and can be used to optimise future surveys (see \cite{f4c} for a detailed discussion on the Fisher Matrix formalism). \name{}~was developed with the aim of providing the community with a free, standard and tested tool for Fisher Matrix analysis, that is both easy to use through the Graphical User Interface, and yet also a robust general base-code for research. The underlying modular code of \name{}~is completely general and is not specific to cosmology although the default setup for the GUI is intended for cosmology. It provides parameter error forecasts for cosmological surveys providing distance, Hubble expansion and growth measurements in a general, curved, FLRW background.

The code is written in Matlab and can be redistributed and modified under the Berkley Software Distribution (BSD) license \cite{BSD} which is now the required license for the Matlab File Exchange\footnote{http://www.mathworks.com/matlabcentral/fileexchange/}, the primary repository for \name. Both the command line and GUI versions of the code produce plots that are generated directly from the program and can be easily edited in the GUI and saved in a variety of standard formats (`.eps', `.jpg', `.pdf' etc...) for inclusion in research publications. There is also a report extension facility which generates either a text or \LaTeX{}~report complete with syntax for incorporating the input and output in tables, matrices and figures. This means that both input and output data and results are easily portable from \name{}~into a research publication.

The simple start-up procedure and ease of use of the \name{}~suite make it well-suited to both teaching and research. The input to the \name{}~code can easily be changed and adapted, hence is can be run in large loops to explore parameter spaces, and for visualisation of the Fisher Matrix.
We now describe the Fisher Matrix framework, outline the start-up procedure of \name{}, and describe the various functions and routines. A shortened version of the start-up procedure of \name{}~is found in the \textbf{Quickstart.pdf} guide -- which is included both as an appendix in \cite{f4c} and in the bundle of \name~software.

\section{Getting Started\label{start}}
Currently the code is available for download at one of the following websites \cite{cosmo_org,mathworks}. Save this `.zip' file into the directory you want to run the Fisher4Cast suite from.

\subsection{The Graphical User Interface}
The GUI can be started from the Matlab editor. The file {\bf FM\_GUI.m} must be opened from the directory, and once the file is opened (click on the file icon from within the Command-line
interface to open it with an editor) press `F5' to run the code. This will open up the GUI screen.

You can also launch the GUI from the command line by typing: \begin{verbatim} >>FM_GUI\end{verbatim} The output data will not be saved into the workspace, but the `Saving Features' button allows one to save the input and output from any particular run in text or \LaTeX{}~code.

\subsubsection{The Basic Layout Explained}
We describe the basic layout of the GUI, and illustrate the various actions with screenshots taken of a working GUI.

\begin{figure}[htbp!]
\begin{center}
\includegraphics[width=.7\textwidth]{\fig_dir{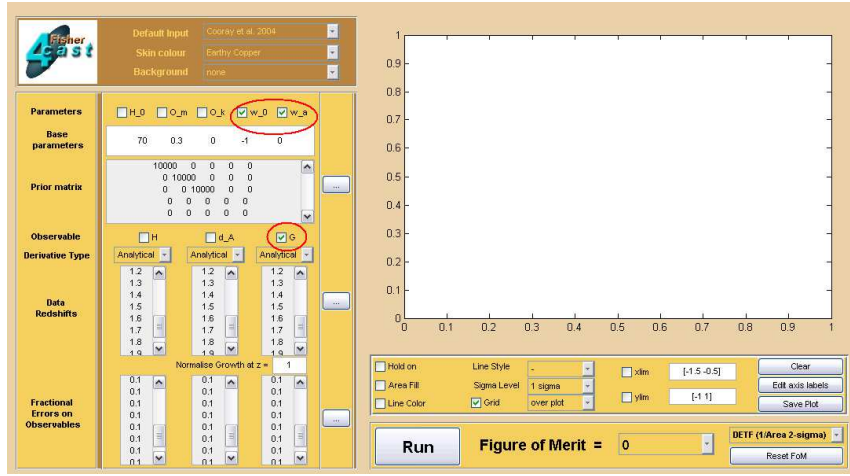}}
\caption{\label{1}The start-up screen of the \name{}~GUI.}
\end{center}
\end{figure}

The GUI has three main sections. The section on the top left controls the input to the GUI. The bottom left panel controls the things one might like to use in the analysis and the parameters you are interested in plotting. In Figure~(\ref{1}) we show the initial GUI screen, highlighting the observable about to be used (here the Growth function $G(z)$), and the cosmological parameters relevant to the analysis (which are the $w_0$ and $w_a$ coefficients in the Chevallier-Polarski-Linder parameterisation of dark energy \cite{cp,lin} -- see Eq.~(\ref{wparam})). The specific cosmological example is described in detail in \cite{f4c}, which contains the set of analytical derivatives used in \name. The right-hand side of the GUI controls the plotting commands for the ellipse. The various actions used to control the output are described below.

\subsubsection{Changing the Input Structure}
\begin{figure}[htbp!]
\begin{center}
\includegraphics[width=.7\textwidth]{\fig_dir{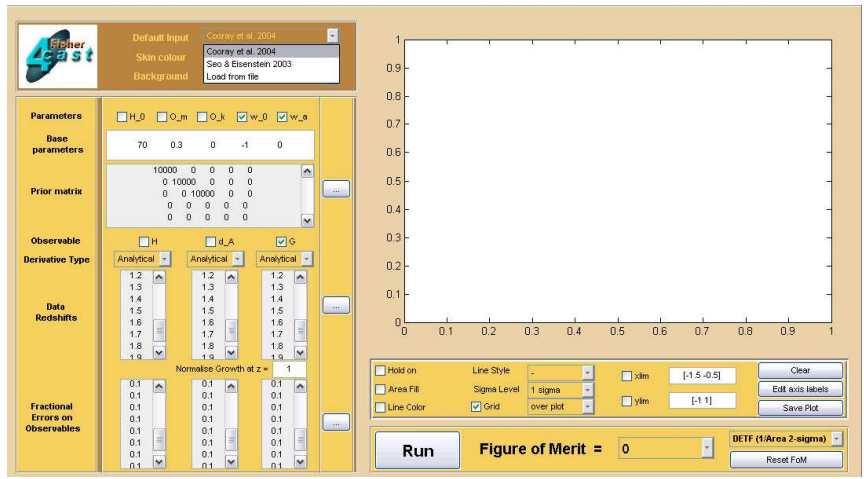}}
\caption{\label{2} Changing the default input structure from the drop-down menu.}
\end{center}
\end{figure}

In order to compute Fisher ellipses for different input structures, one can either choose from a drop down list of default example structures contained within the distribution (as shown in Figure~(\ref{2})) or one can generate a unique input structure. This file can then be loaded to the GUI, which must be given as `.m' file. You can obviously also just edit the input parameters in the GUI after the default input has been loaded or alternately you can edit the input file (eg {\bf Cooray\_et\_al\_2004.m}).

\subsubsection{Floating Help}
\begin{figure}[htbp!]
\begin{center}
\includegraphics[width=.7\textwidth]{\fig_dir{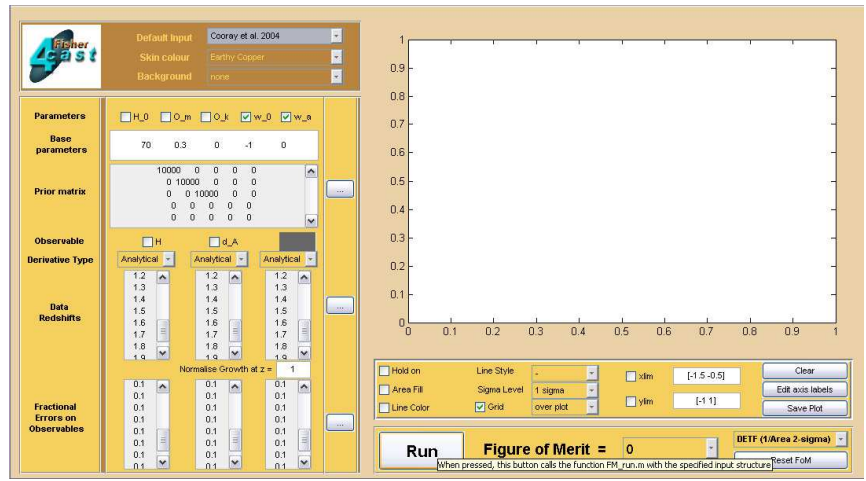}}
\caption{\label{3} The floating help for the `Run' button.}
\end{center}
\end{figure}

Floating help is provided with the \name{}~GUI for most commands. The floating help is activated by moving the mouse pointer over the button or parameter on the GUI and leaving it there for a few seconds. This generates a screen prompt, which pops up and gives information about the function of the button or parameter in question. Figure~(\ref{3}) shows this help prompt for the `Run' button on the GUI.

\subsubsection{Running \name{}}
\begin{figure}[htbp!]
\begin{center}
\includegraphics[width=.7\textwidth]{\fig_dir{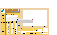}}
\caption{\label{4} Running the \name{}~from the GUI.}
\end{center}
\end{figure}

Once satisfied with the observables considered and the parameters of interest, pressing the `Run' button will execute the code. A box will pop up that will state that the code is running, and an error ellipse will appear when the code has finished running. This is shown in Figure~(\ref{4}).

\subsubsection{Errors in the Input}
\begin{figure}[htbp!]
\begin{center}
\includegraphics[width=.7\textwidth]{\fig_dir{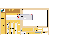}}
\caption{\label{5} The GUI error message box. The errors are detailed in `log.mat'.}
\end{center}
\end{figure}

When the `Run' button is pushed, the GUI first calls the {\bf FM\_errorchecker.m} function with the input supplied. This checks for the input files, checks that the data vector (e.g. redshifts at which one has measurements of the Hubble parameter) and error vectors (e.g. the fractional errors on the Hubble parameter, $\sigma_H/H,$ at the redshifts above) are the same length and performs other consistency checks. Should any of these tests fail, an error box will appear explaining which errors to fix before calling the GUI again. A log file of these errors is created in the same directory the GUI is being run in, and is called `log.mat'. Loading and reading this log file is described in Section~\ref{errorchecker}. Figure~(\ref{5}) shows the error dialogue box indicating that a single error has been found.

\subsubsection{The Fisher Ellipse}
\begin{figure}[htbp!]
\begin{center}
\includegraphics[width=.7\textwidth]{\fig_dir{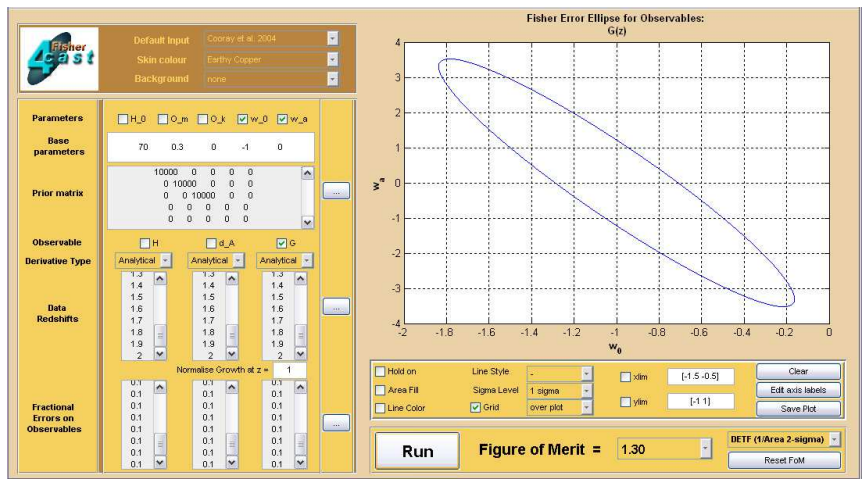}}
\caption{\label{6} The Fisher ellipse resulting from a run of \name.}
\end{center}
\end{figure}

Once the code is running smoothly, the resulting Fisher error ellipse is plotted. This is shown in Figure~(\ref{6}).

\subsubsection{Plotting more than one Ellipse}
\begin{figure}[htbp!]
\begin{center}
\includegraphics[width=.7\textwidth]{\fig_dir{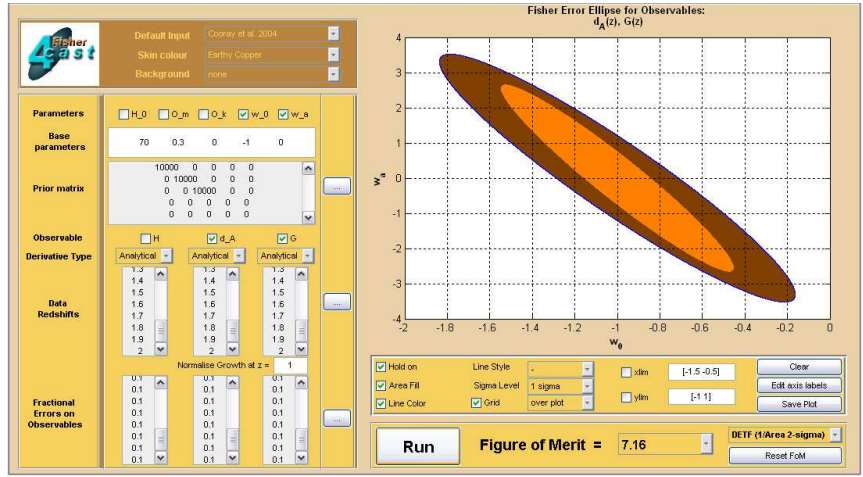}}
\caption{\label{8} Multiple ellipses are plotted in one figure.}
\end{center}
\end{figure}

Should one want to superimpose more than one ellipse, click the `Hold on' button. This works both for the line and the area (although the same line and area fill properties will be used for both ellipses -- see the below item for discussion of changing the colour of the area fill). Figure~(\ref{8}) shows the resulting ellipse for two observables, $G(z), d_A(z)$.

\subsubsection{Area fill}
\begin{figure}[htbp!]
\begin{center}
\includegraphics[width=.7\textwidth]{\fig_dir{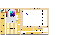}}
\caption{\label{7} The Area Fill option with colour selection.}
\end{center}
\end{figure}

Clicking on this button yields a filled error ellipse. Once it is clicked a colour must be selected from the menu on the left pop-up box. Note that should more than one error ellipse be plotted later, this area fill box must be ticked and un-ticked again to change the colour, otherwise the same colour will be used for all filled ellipses. This box is shown in Figure~(\ref{7}).

\subsubsection{Importing Data}
\begin{figure}[htbp!]
\begin{center}
\includegraphics[width=.7\textwidth]{\fig_dir{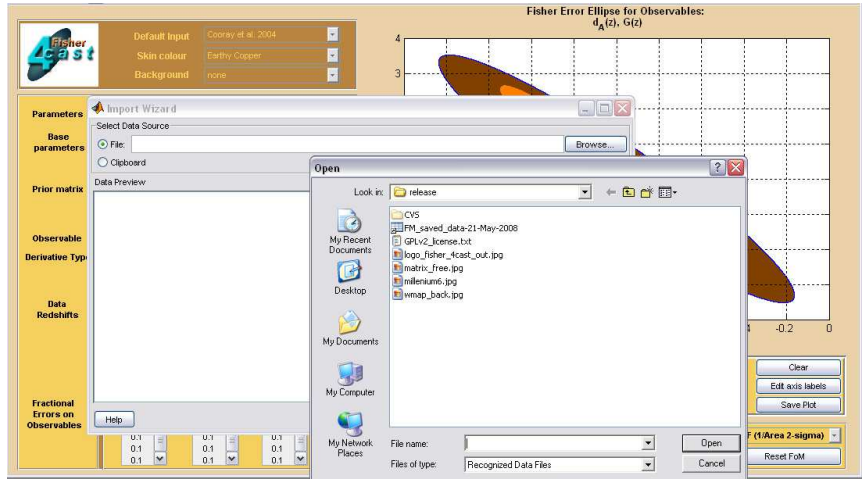}}
\caption{\label{9} Loading data into the GUI.}
\end{center}
\end{figure}

The input data can also be imported from a file - either as the redshift vector (the data), the error vector or the matrix of prior information on the cosmological parameters. This can be done by clicking the relevant `Browse' buttons on the GUI. This brings up a screen in which one can either load the data from file, or from the clipboard, in which case the data is cut and paste into the GUI fields. Figure~(\ref{9}) shows the screens for the loading of data from a file in the directory. In addition there is a check-box which specifies whether or not to use the prior matrix.
\subsubsection{Multiple $\sigma$}
\begin{figure}[htbp!]
\begin{center}
\includegraphics[width=.7\textwidth]{\fig_dir{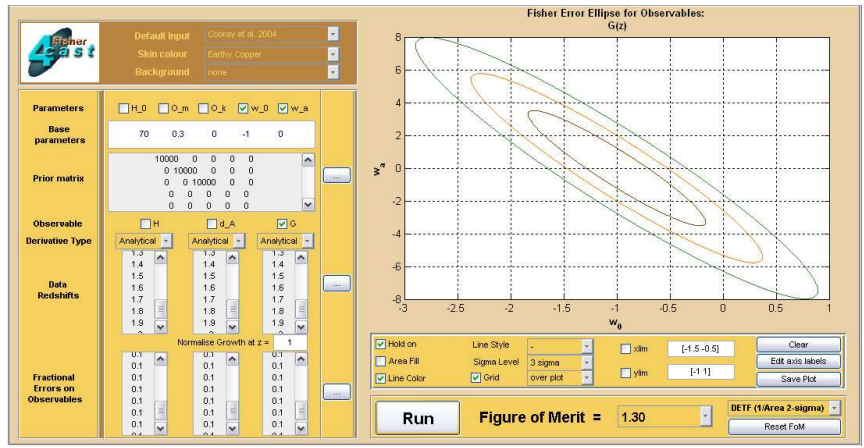}}
\caption{\label{10}Ellipses with many $\sigma$.}
\end{center}
\end{figure}

It is possible to plot the ellipses for multiple confidence levels (i.e. $67\%, 95\%, 99\%$ specified by $1-, 2-,~\mathrm{and}~3-\sigma$ respectively). This is done via a drop-down menu on the right-hand side of the GUI, and is illustrated in Figure~(\ref{10}). It is worth noting that the `Hold on' was used to generate this plot.

\subsubsection{Different Definitions of the Figure of Merit}
\name{}~provides various Figures of Merit in order to compare different surveys. These are defined in detail in Section~\ref{sec:FM_fom}. Figure~(\ref{13}) illustrates how these various FoMs are accessible in the GUI.
\begin{figure}[htbp!]
\begin{center}
\includegraphics[width=.7\textwidth]{\fig_dir{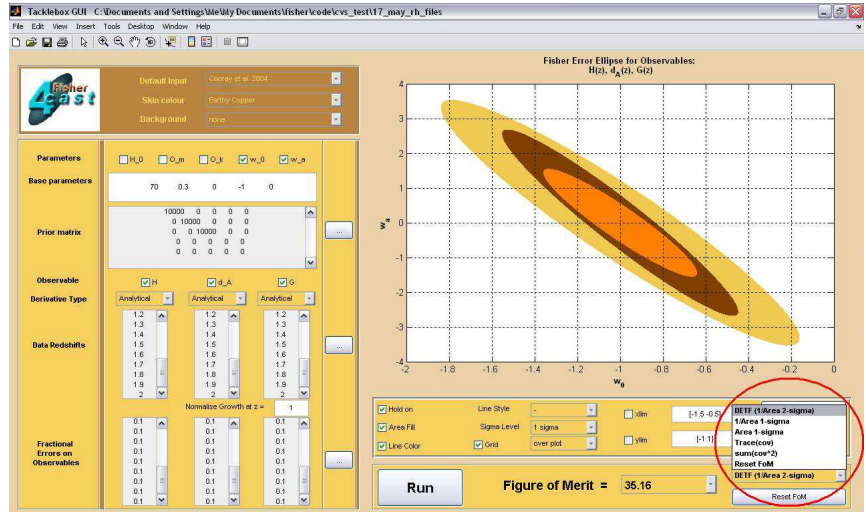}}
\caption{\label{13}Various Figures of Merit $\sigma$ are calculated in \name.}
\end{center}
\end{figure}
\subsubsection{Controlling Output}
The buttons on the right-hand side of the GUI all control the output specifications of the ellipse, such as the limits of the $x$ and $y$ axis, the line style and colour of the ellipse, and whether or not to have a grid on (over or under) the data. This is designed for maximum flexibility in representing the ellipses in a unique and distinguishable way. The axis labels can also be modified using the `Edit Axis' button.
\subsubsection{Saving the Plot}
\begin{figure}[htbp!]
\begin{center}
\includegraphics[width=.7\textwidth]{\fig_dir{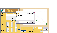}}
\caption{\label{11}Saving the figure of your \name~ellipse.}
\end{center}
\end{figure}

Once satisfied with the ellipses plotted, figures are saved by clicking on the `Saving Features' menu and selecting the 'Save Plot' option. This will bring up a window to save the figure to a particular directory, in a selected file format (`.eps', `.fig', `.png', `.pdf' etc.), as illustrated in Figure~(\ref{11}). No legend information (for example which colours correspond to which ellipses) will be saved. If legend entries are required, the figure must be saved as a `.fig' file, opened with Matlab, with the required legend entries added later.
In addition to the simple saving of the figure, the reporting features of the \name{}~code are accessed through this `Saving Features' menu. In short, they provide the opportunity to save a report of the inputs and the resulting outputs and Fisher matrices from the survey, in either ASCII text or \LaTeX{}~format. When clicking the `Text report' feature the user is prompted to save the resulting `.txt' file. Similarly if one chooses the \LaTeX{}~report, both a `.tex' report and an Encapsulated Postscript File~will be generated, and the user is prompted to save both the figure and the \LaTeX{}~document. These reporting features are discussed in more detail in Section~\ref{report}.
\subsubsection{Skins}
\begin{figure}[htbp!]
\begin{center}
\includegraphics[width=.7\textwidth]{\fig_dir{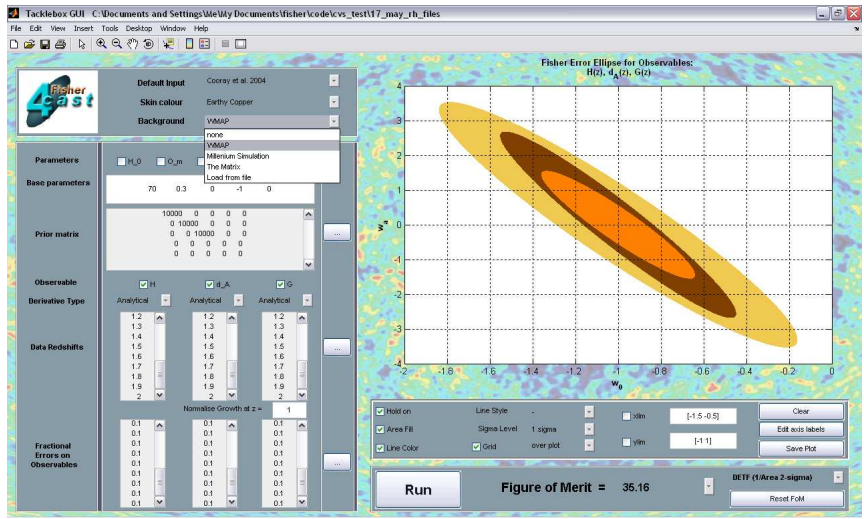}}
\caption{\label{12}Changing the skins and background images.}
\end{center}
\end{figure}

The GUI is available in a variety of skins and backgrounds. These can be chosen from a drop-down list (consisting of both colour schemes and background images \cite{millenium_sim, wmap_back,matrix_free}); additional background images can be loaded by the user. This is shown in Figure~(\ref{12}).

\subsubsection{Fisher4Cast Menu}
A \name{}~menu is defined in the top left-hand corner of the \name{}~GUI. From this drop-down menu one can access the {\bf Readme} file of the code suite, the {\bf Users' manual} and {\bf Quickstart Guide} for easy reference, and the version history of the code. The BSD licence \cite{BSD} for the \name{}~suite is also available from the drop-down list. This list is illustrated in Figure~(\ref{13}). In addition to the \name{}~menu, there is a menu with information on the extensions available for \name{}. In future releases of the code this menu will select the extensions themselves, at present it provides the {\bf Readme} for the modules.

\begin{figure}[htbp!]
\begin{center}
\includegraphics[width=4.7in]{\fig_dir{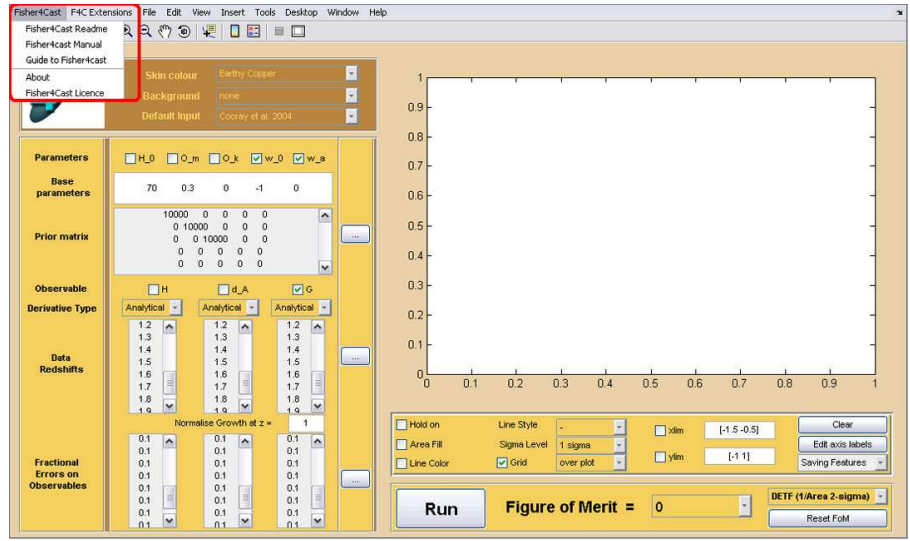}}
\caption{\label{13}The \name{}~drop-down menu with information on the code and version.}
\end{center}
\end{figure}
\subsubsection{Interactive Plotting}
Interactive `point-and-click' plotting is available in \name~\versionnum, available by selecting the `Activate Interactive Plotting' option from the `Fisher4Cast Extension' menu. Once selected the user interactively sets the values for parameters being plotted by clicking on the plotting area of the GUI. The arrow is activated for the first click on the plot area, the next click will run the code and produce the appropriate ellipse. The values selected will be displayed in the parameter input section of the GUI, the same as they would have been if manually entered. Care should be taken not to step to very unphysical values of the parameters, e.g. very positive values of $w_0$.
\begin{figure}[htbp!]
\begin{center}
\includegraphics[width=.7\textwidth]{\fig_dir{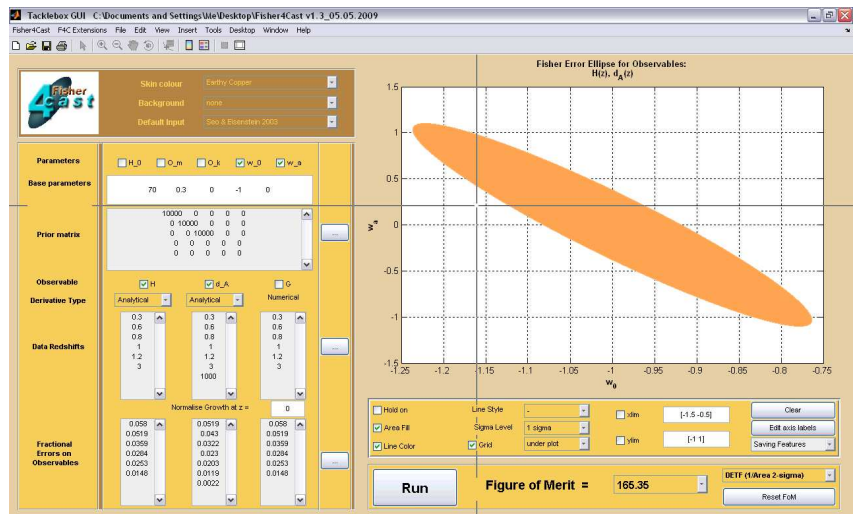}}
\caption{\label{13}The Interactive Plotting Feature in the \name{}~code. This allows you to click on the parameter plane to change the assumed fiducial model.}
\end{center}
\end{figure}

\subsection{The Command Line}

\subsubsection{Running the Code}
Open your version of Matlab and change the working directory to be the same as where you saved \name{}~in. To run the code from the command line with one of the standard test input structures supplied, type:\begin{verbatim} >>output = FM_run(Cooray_et_al_2004)\end{verbatim}

This will call the code using the pre-supplied test input data ({\bf Cooray\_et\_al\_2004}) and then generate an error ellipse plot for the parameters and observables supplied in the chosen input. All the relevant generated output is written to the output structure. You can see the range of outputs to access by typing:\begin{verbatim} >>output \end{verbatim}
and then examine each output individually by specifying it exactly.
For example: \begin{verbatim} >>output.marginalised_matrix\end{verbatim} will access the marginalised\_matrix field in the output structure. It is worth noting that each `.' denotes another sub-level in the input structure.

Example input files are supplied as a template for generating new input files with your own customised parameters and values. All fields specified in the example inputs must be specified in any user-defined example input. These are outlined in Section~\ref{init}.

The code can also be run from the Matlab editor. Once the code is opened (open it from inside the Matlab window), pressing `F5' will run the code. Note that if the code is run from the Editor it will call the default input structure, which is the {\bf Cooray\_et\_al\_2004.m} file. This is an example file containing input data from the paper by Cooray {\em et al.} \cite{cooray2004}. This output can be directly compared to that of Figure~1 of that paper. If your output compares correctly, you have a working installation of the code. Another input available is {\bf Seo\_Eisenstein\_2003.m }\cite{seo2003}.
\subsection{FM\_errorchecker}
\label{errorchecker}
The error-checker function acts `behind-the-scenes' to check that the input structure and all the required variables are correct before executing the code. It can be run directly by using the command:
\begin{verbatim}
 >>FM_errorcheck(FM_initialise)
\end{verbatim} where {\bf FM\_initialise} is the specific function to initialise the input structure.
The error checker validates, among other things, that all the derivative functions (whether analytical or numerical derivatives are going to be implemented) do in fact exist and that the data and corresponding variances vectors are the same length. This error checker is continually being updated to facilitate ease of use of the code. All error and checking messages are displayed to the screen and are also saved in the Matlab file `log.mat' which can be loaded and examined at a later stage by invoking the following command:
\begin{verbatim}
 >>load(log.mat)
\end{verbatim}

\section{Flowchart \label{flowchapt}}
Throughout the code structures are used to allow different sub-parts of the general code access to the data. These structures are defined as global variables. The structures containing information on the input for the code are either defined at the beginning or loaded from file. All output structures can be saved for later use. We now outline the general framework of the code and describe the cosmological example specifically coded in \name{}.
\begin{figure}[htbp!]
\begin{center}
\includegraphics[width=3.9in]{\fig_dir{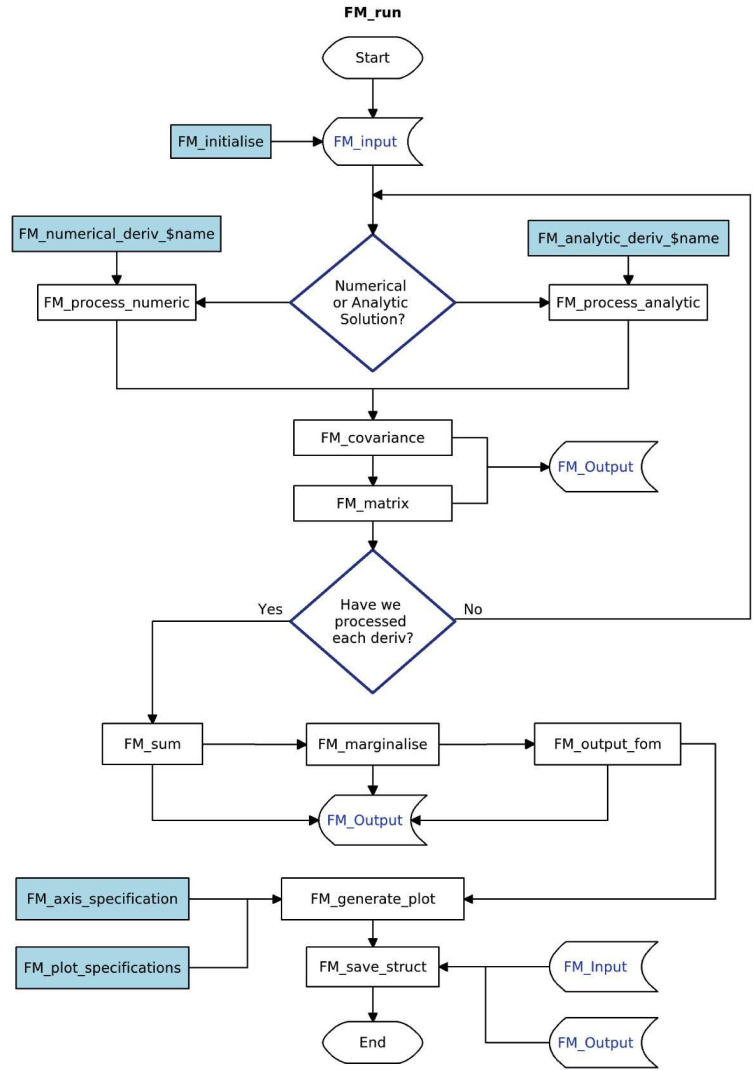}}
\caption{\label{ft}The flowchart of the code showing the processes, decisions and storage of data. For a key to the symbols see Figure~(\ref{ft_key}).}
\end{center}
\end{figure}
The flowchart shown in Figure~(\ref{ft}) summarises the layout and structure of the code, and the symbols are summarised in Figure~(\ref{ft_key}).
\begin{figure}
\centering
\includegraphics[width=0.7\textwidth]{\fig_dir{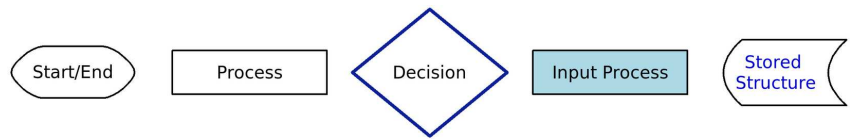}}
\caption{\label{ft_key}This key shows the symbols contained in the flowchart, Figure~(\ref{ft}). They are from, left to right, the begin and terminate indicator; a simple processing function which would generally return an output; an if statement or for loop; an input process function designed to be edited and changed as per the user specifications and lastly a stored structure for either input or output and passed globally for use throughout the code.}
\end{figure}
\newpage
\section{Components of the Code \label{compchapt}}
We now discuss the various components of the code in detail. With each section we give a subsection of the flowchart to highlight the position in the flow of information through the code structure.
\subsection{FM\_run}
\begin{figure}[htbp!]
\begin{flushleft}
\includegraphics[width=2cm]{\fig_dir{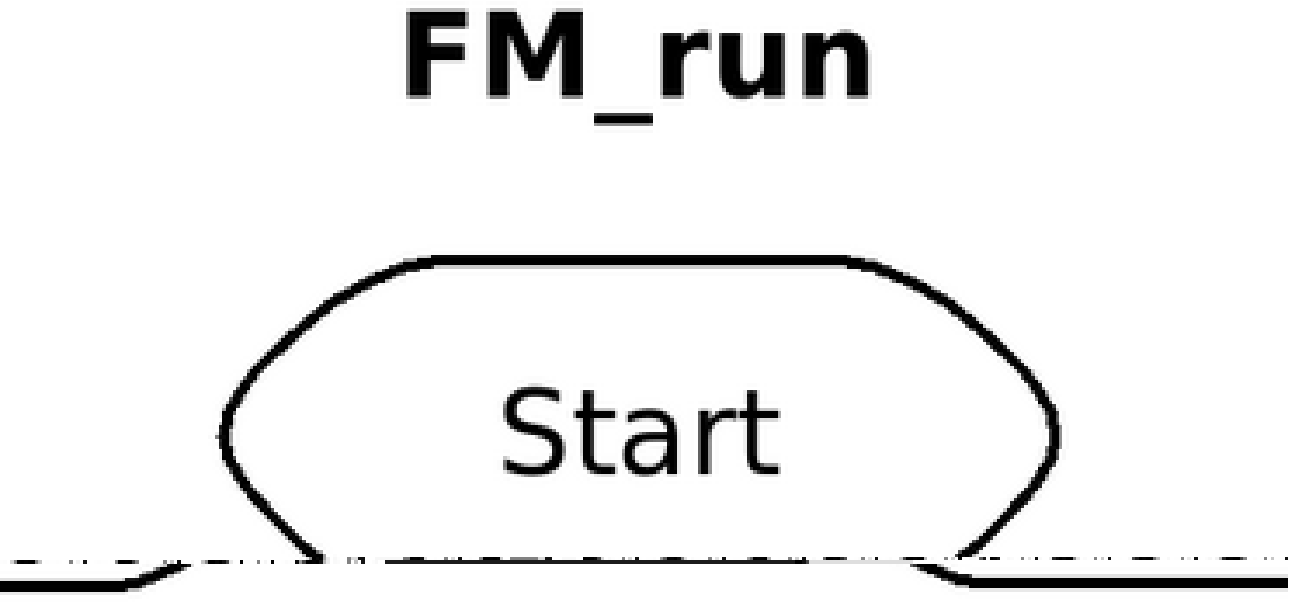}}
\end{flushleft}
\end{figure}
{\bf FM\_run.m} is the general wrapper of the code. In order to make the code clear and easily editable, all main processes are called from this general function, and it is where all data storage occurs. Links to separate functions for specific calculations are documented in the code. From the command line this code is called with one argument, namely the specific function that initialises the input structure. This is unique to each example. As outlined in the section on implementing the code, if no argument is given the code is run with a pre-defined function, given by {\bf Cooray\_et\_al\_2004.m}, which gives the parameters for a redshift survey as outlined by Cooray {\em et al.} \cite{cooray2004}.

\subsection{FM\_initialise \label{init}}
\begin{figure}[htbp!]
\begin{flushleft}
  \includegraphics[width=4.5cm]{\fig_dir{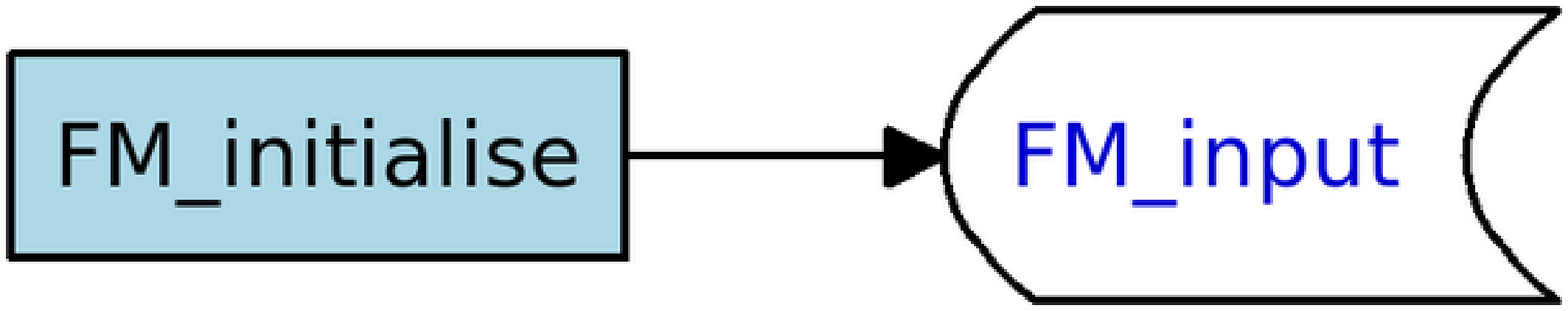}}
\end{flushleft}
\end{figure}
This function initialises the input used throughout the processing of the \name{}~code. The values, names and areas of interest are specified here. It is called by {\bf FM\_run.m} to set the initial values for the {\bf input} structure, which is then passed globally to all other parts of the code. Examples of the default initialising functions provided are {\bf Cooray\_et\_al\_2004.m, Seo\_Eisenstein\_2003.m} \cite{cooray2004,seo2003}. It is important to note that the format of the initialising function must be kept constant - in other words the same input must be specified in each initialising function - the code expects values for certain entries in the {\bf input} structure. The entries are as follows:
\begin{itemize}
\item
{\bf input.function\_names} - A cell of strings containing the specific filenames of the {\em analytical derivatives}. Note in the coded cosmological example that no analytical derivative function is specified for the growth function, derivatives are only taken numerically.
\item
{\bf input.observable\_names} - A cell of strings with the names of the observables.
\item
{\bf input.observable\_index} - A vector of the indices corresponding to the observable\_names you are interested in, eg [2 3] would imply that your considering the second and third observables listed in observable\_names.
\item
{\bf input.data$\{$i$\}$} -  These are the row vectors of the data for each of the respective observables (indexed again from beginning to end by $i$). In the cosmological example {\bf input.data$\{$1$\}$} would be the redshifts at which you have measurements of the Hubble parameter, for example.
\item
{\bf input.parameter\_names} - A cell of strings containing the names of the parameters you can include for consideration to generate Fisher Ellipses.
\item
{\bf input.base\_parameters} - A row vector of the parameter values (they must be specified with the same order as the parameter\_names vector). This is the model assumed to be true in the analysis, the Fisher Matrix is taken around this fiducial model.
\item
{\bf input.prior\_matrix} - The prior matrix for the parameters taken from previous surveys etc. The order of the matrix columns and rows correspond to the respective parameters listed as your parameter\_names.
\item
{\bf input.parameters\_to\_plot} -  A row vector of the indices of the specific parameters you want to plot. If one index is given then a likelihood function for that parameter is plotted and if two are specified then an error ellipses is plotted. Selecting more than two parameters will produce an error message, as \name{}~is only coded for up to 2-dimensional error contours.
\item
{\bf input.num\_parameters} - This is a derived value and is given by the length of the number of parameters you are considering in {\bf parameters\_to\_plot}.
\item
{\bf input.num\_observables} - This is a derived value and is generated from the number of observables under consideration in observable\_index.
\item
{\bf input.error\{i\}} - The fractional error on the data from the observables ($\sigma_{\bf X^\alpha}/{\bf X^\alpha}$). It is key that there are as many error entries as there are observables you are considering (i.e. {\bf input.error$\{$1$\}$} gives the error on your measurements of the Hubble parameter, measured at {\bf input.data$\{$1$\}$}). The entry can either be a row vector, in the case of uncorrelated observables (this vector is converted to a diagonal matrix in the code) or a covariance matrix.
\item
{\bf input.numderiv.flag} - A logical entry is expected here for each observable, should you wish to use numerical derivatives.
\item
{\bf input.numderiv.f} - Single string entries which are combined into a struct later. These entries are only required if you have specified that you would like numerical derivatives for your observables. They give the function name of the function (say {\bf g.m}) of which you are taking derivatives.
\end{itemize}

\subsection{The Derivative Loop \label{deriv_loop}}
\begin{figure}[htbp!]
\begin{flushleft}
\includegraphics[width=9.5cm]{\fig_dir{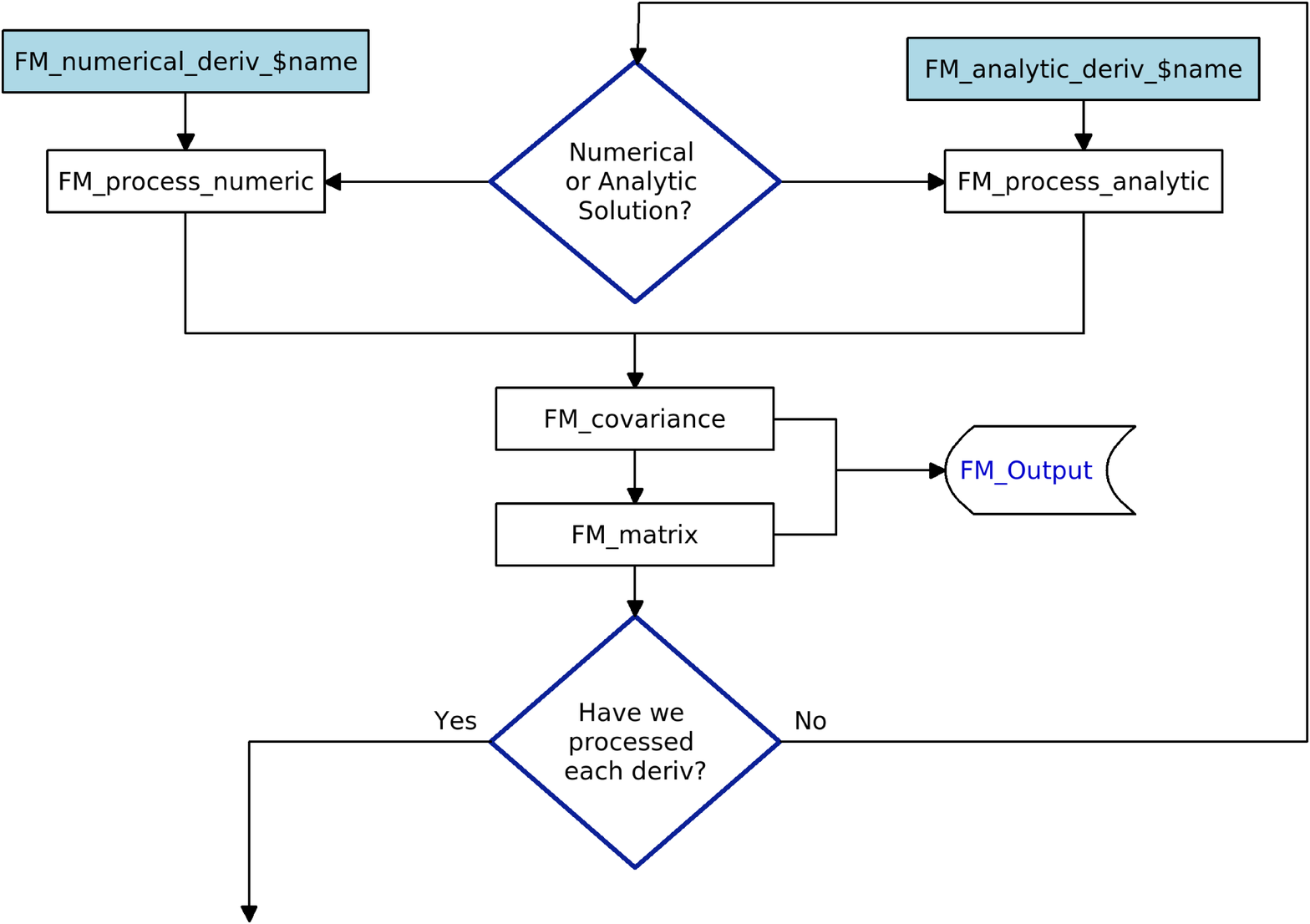}}
\end{flushleft}
\end{figure}
The code now runs various operations in a loop over the specific observables. For each observable it checks whether numerical or analytical derivatives are to be used by checking the numerical flag as specified in the input structure (see the above discussion on the numderiv.flag for the input structure). Both the analytical and numerical derivatives return a matrix of derivatives for all the parameters and observables as well as a vector of the function evaluated at the data points specified. The specific details of the numerical and analytical derivative codes are outlined in the proceeding discussion. Once the selected derivative process is completed, the relevant output is stored in the {\bf output} structure.

\subsubsection{Numerical Derivatives}
\begin{figure}[htbp!]
\begin{flushleft}
\includegraphics[width=3cm]{\fig_dir{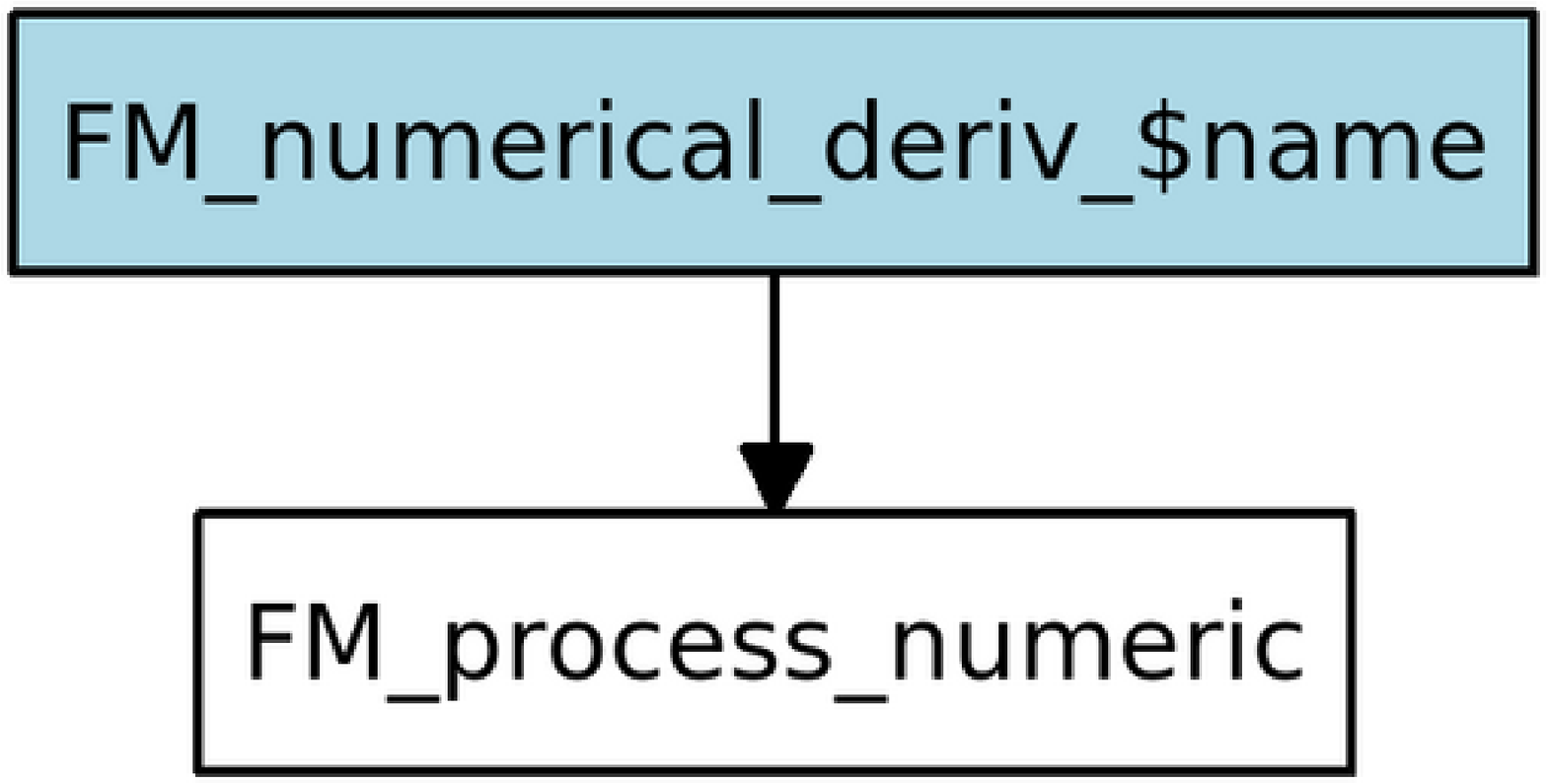}}
\end{flushleft}
\end{figure}
The numerical derivative code will calculate the numerical derivatives of any function (say {\bf g.m}) provided that the function is specified as a function of the input parameters (i.e. $g = g({\bf d}, \theta_\mathrm{A}, \theta_\mathrm{B},...)$), by calling on {\bf FM\_process\_numeric.m} which in turn calls {\bf FM\_num\_deriv.m} and passes it the name of the function you wish to take derivatives of.

The standard numerical derivative algorithm used is known as the complex-step method \cite{complex_step}: \[\frac{\partial g }{\partial \theta_\mathrm{A}} = \mathrm{Im}\left[\frac{g({\bf d},\theta_\mathrm{A} + ih, \theta_\mathrm{B},...)}{h}\right], \] where $\mathrm{Im}$ represents the imaginary part of the argument, and $i^2 = -1$ as usual. This method is a second order accurate formula and is not subject to subtractive cancellation. Unlike the finite-difference method an arbitrarily small step-size can be chosen and therefore the complex-step method can achieve near analytical accuracy.

In addition, the simple double-sided central derivative is coded in the {\bf FM\_num\_deriv.m} function. In order to use this algorithm the user must change the method field inside the derivative function from `complex'  to `central'.  In this case the gradient is then calculated as \[\frac{\partial g}{\partial \theta_{\mathrm{A}}} = \frac{g({\bf d},\theta_\mathrm{A} + h, \theta_{\mathrm{B}},...) - g({\bf d},\theta_{\mathrm{A}} - h, \theta_\mathrm{B},...)}{2 h}. \] This is then iterated until the gradient converges for the parameter. Note that the convergence criterion is quite stringent - and an error message will result if there are possible convergence issues. However this criterion can be relaxed by changing the settings in the {\bf FM\_num\_deriv.m} code.

Once the derivatives are saved the Fisher Matrix must be calculated for this observable. This is done by first calculating the data covariance matrix for the observable. This is done in {\bf FM\_covariance.m} which is passed the function value and the index of the observable. The code checks if the error entry is a covariance matrix (in the case of correlated observables) or a vector in the uncorrelated case. It then calculates the covariance matrix by multiplying the variance with the function value at the data points considered.

{\bf FM\_matrix.m} then produces a Fisher Matrix ($F$) from the covariance matrix ($C$) and the derivative matrix ($V$) using matrix multiplication as $F = V^{\mathrm{T}}C^{-1}V $.
\subsubsection{Analytical Derivatives}
\begin{figure}[htbp!]
\begin{flushleft}
 \includegraphics[width=2.8cm]{\fig_dir{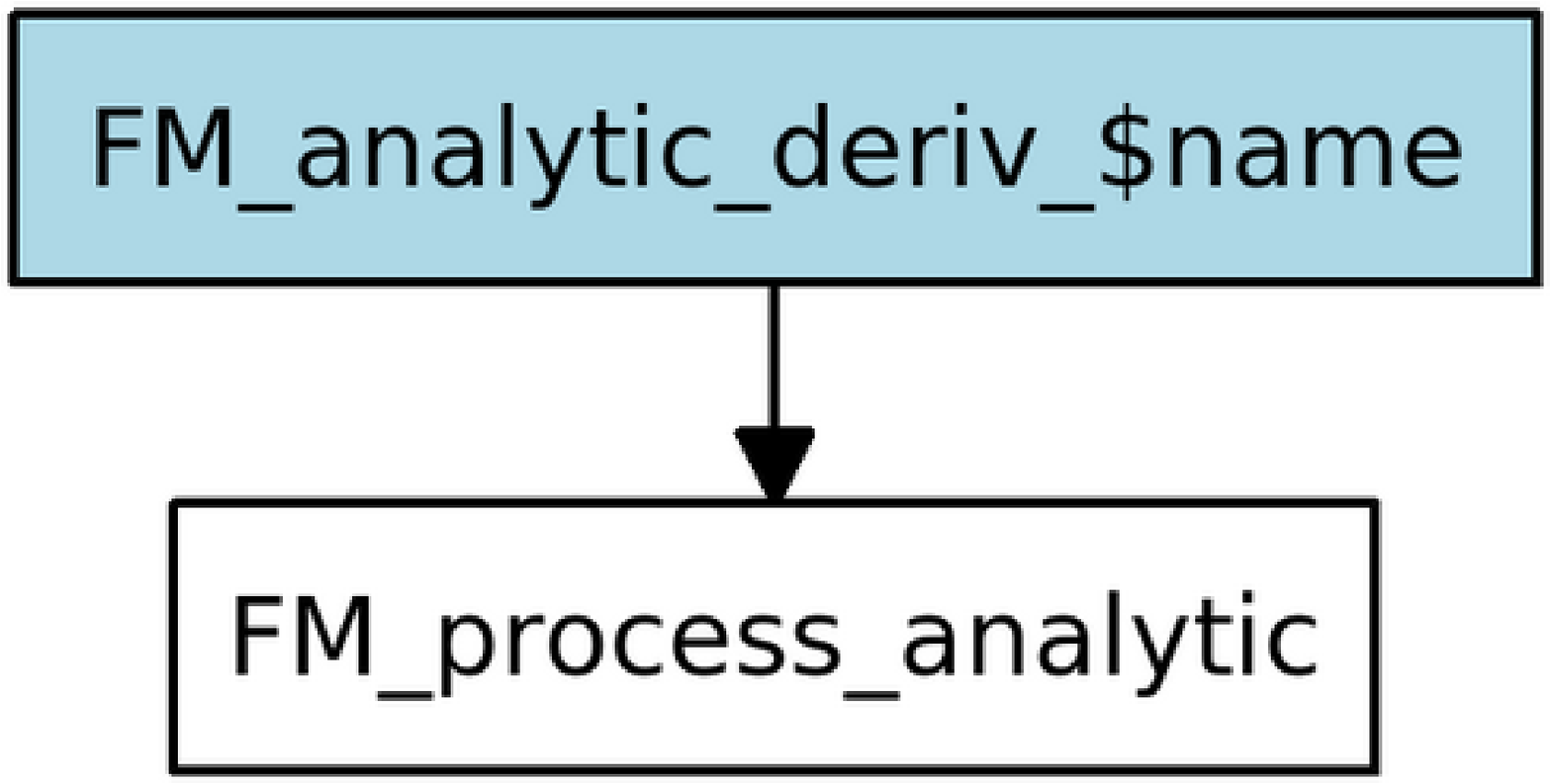}}
\end{flushleft}
\end{figure}

The analytical derivatives are specific to each user and example. If one knows the analytical form of both the function and of the Fisher derivatives, one can include these functions explicitly. The only conditions on these functions are that they must be of the form $g = d({\bf d}, {\boldsymbol \theta})$ and must return as output a matrix of Fisher derivatives $\partial g/\partial \boldsymbol \theta$ and a vector of the function itself evaluated at the data points $\bf d$ given in the {\bf input.data$\{$i$\}
$}. These derivative functions are supplied for the Hubble parameter and angular diameter distance as {\bf FM\_analytic\_deriv\_1.m} and {\bf FM\_analytic\_deriv\_2.m} respectively. The Fisher derivatives of the angular diameter distance with respect to the cosmological parameter $\Omega_k$ must be taken as Taylor series expansion when $\Omega_k \rightarrow 0$ (see \cite{f4c} for the full set of derivatives in \name).

As in the numerical derivative case, once the derivatives are saved the Fisher Matrix must be calculated for this observable. This is done by calculating the data covariance matrix for the observable in {\bf FM\_covariance.m} which is passed the function value and the index $\alpha$ of the observable. The code checks if the errors are specified a covariance matrix (in the case of correlated observables) or a vector in the uncorrelated case. It then calculates the covariance matrix by multiplying the variance with the function value at the data points considered.

{\bf FM\_matrix.m} then produces a Fisher Matrix ($F$) from the covariance matrix ($C$) and the derivative matrix ($V$) using matrix multiplication as $ F = V^{\mathrm{T}}C^{-1}V $.

\subsection{Final processing}
\label{sec:FM_fom}
\begin{figure}[htbp!]
\begin{flushleft}
    \includegraphics[width=8cm]{\fig_dir{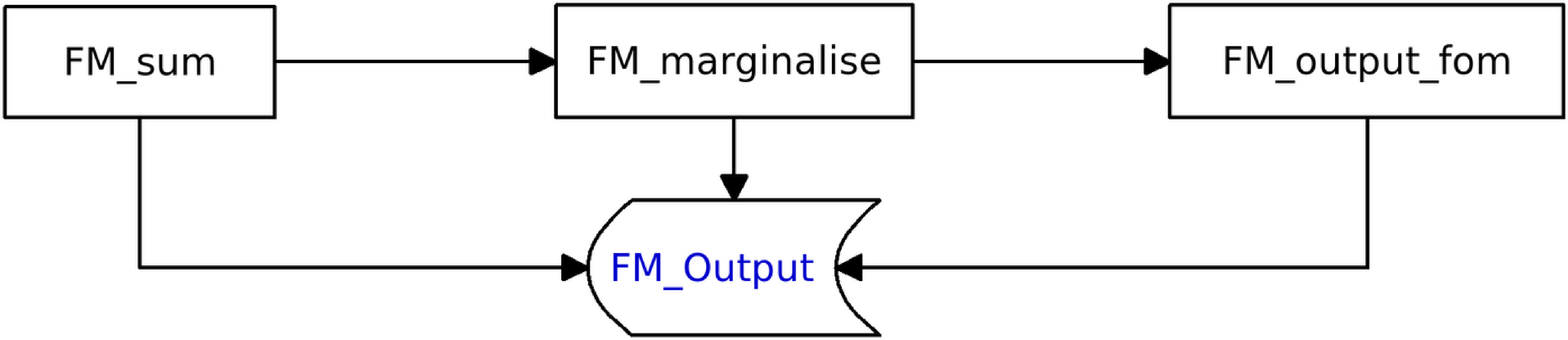}}
\end{flushleft}
\end{figure}

{\bf FM\_sum.m} collates all the derivative matrices from the previous steps and sums them together to form a full Fisher Matrix. The individual Fisher matrices for each observable are added to the prior matrix (as specified in the {\bf input} structure). This complete Fisher Matrix is general and is assigned to the {\bf output} structure for future reference.

{\bf FM\_marginalise.m} produces a marginalised Fisher Matrix (say $\widetilde{F}$). It takes the parameters you are interested in (specified as {\bf parameters\_to\_plot} in the {\bf input } structure) and shuffles the Fisher Matrix into a block form. It then performs matrix multiplication on the blocks to produced the marginalised Fisher Matrix, which is also assigned to the output structure.

{\bf FM\_output\_fom} then produces the appropriate error for the likelihood case and a range of FoMs, listed below, for the case of an ellipse (e.g. if the length of {\bf parameters\_to\_plot} in the {\bf input} structure is two then an ellipse will be plotted). These Figures of Merit are then stored in the {\bf output} structure. \name{}~includes the standard FoMs as well as some new ones available through the GUI and command line. Although some of the Figures of Merit are only defined for the error ellipse in the $w_0-w_a$ plane, where $w_0, w_a$ are the coefficients in the CPL \cite{cp, linder_w} parameterisation of the equation of state of dark energy (see for e.g. \cite{correct_detf}), the FoMs in \name{}~are calculated by the code for the pair of cosmological parameters being considered rather than the full 5-D matrix. We briefly outline the FoMs used in \name{}:
\begin{itemize}
\item{DETF}\\
This Figure of Merit in the Report of the Dark Energy Task Force \cite{correct_detf} is defined to be the reciprocal of the area of the $2-\sigma$ error ellipse in the $w_0-w_a$ plane of the CPL dark energy parameterisation \cite{cp, linder_w}. This is, equal to $\mathrm{det}(F^{1/2})/(\pi\sqrt{6.17})$. Unfortunately the DETF report does not appear to use this definition, and instead quotes $\mathrm{det}(F^{1/2})$, which is the inverse of the $1-\sigma$ ellipse in units of the area of the unit circle. Because of the benefits of the geometric interpretation \name{}~returns the true inverse area of the $2-\sigma$ ellipse. To convert from one DETF FoM to the other, one should multiply the \name{}~DETF output by $\pi \sqrt{6.17} \simeq 7.8$ 
\item{Area$^{-1}_{1-\sigma}$}\\
This Figure of Merit is the reciprocal of the $1-\sigma$ error ellipse area in the parameter plane currently plotted, i.e. $\mathrm{det}(F^{1/2})/(\pi\sqrt{2.31})$
\item{Area$_{1-\sigma}$}\\
Simply the inverse of the previous FoM.
\item{Tr${\mathbf{C}}$}\\
This FoM is defined as the trace of the covariance matrix of the data, $\mathbf{C} = \mathbf{F^{-1}}$, estimated as the inverse of the marginalised Fisher Matrix. This FoM is simply the sum of the squares of the marginalised errors on each parameter.
\item{$\sum_{\mathrm{AB}} C_{\mathrm{AB}}^2$} \\
This FoM is defined as the sum of the squares of the entries of the whole covariance matrix, $\mathbf{C} = \mathbf{F^{-1}}$. Unlike the previous definition this FoM is sensitive to the off-diagonal components of the covariance matrix as well as the diagonal components.
\end{itemize}

\subsubsection{FM\_output}
\begin{figure}[htbp!]
\begin{flushleft}
\includegraphics[width=1.8cm]{\fig_dir{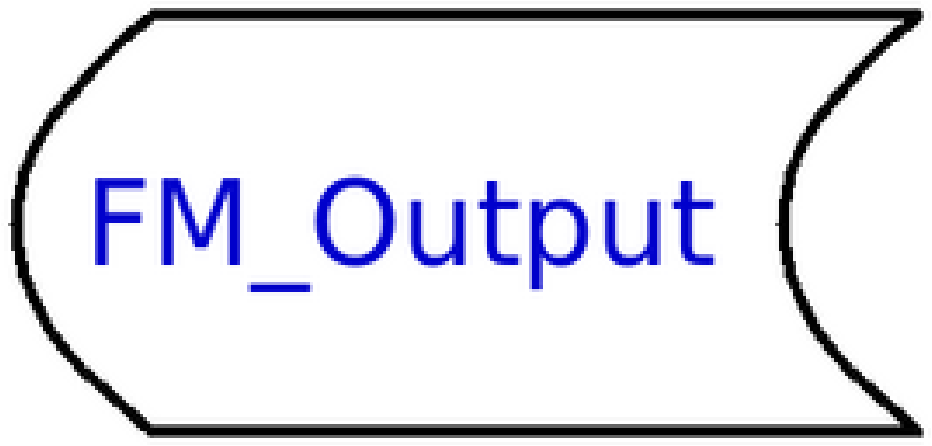}}
\end{flushleft}
\end{figure}
The data from all parts of the code are saved in the {\bf output} structure. The structure formalism in Matlab means that each `.' indicates a further sub-level in the structure. Entries in the structure are of mixed type (i.e. {\bf output.function\_value} is a cell of vectors, one for each observable, while {\bf output.function\_derivative} is a cell of matrices of derivatives, again with one matrix of derivatives for each observable). By the end of the execution of {\bf FM\_run.m} the {\bf output} structure should have the following entries:
\begin{itemize}
\item
{\bf output.function\_value} - This is a cell which contains a set of vectors for each of the observables considered. If in the specific run of the code you have only calculated an error ellipse for say one out of three observables then the rest of the entries are empty vectors.
\item
{\bf output.function\_derivative} - This cell now contains matrices of the Fisher derivatives for the observable. Again, the entries of observables you are not considering will result in empty matrices.
\item
{\bf output.data\_covariance} - This cell contains the calculated data covariance matrix corresponding to each of the observables considered.
\item
{\bf output.matrix} - This cell contains a separate Fisher Matrix for each of the observables considered.
\item
{\bf output.summed\_matrix} - This cell contains the sum of the Fisher matrices for each observable, and the prior information matrix, if included.
\item
{\bf output.marginalised\_matrix} - The marginalised Fisher Matrix given here depends on which parameters are of interest in each run of the code. The marginalisation via matrix multiplication is outlined in Section~\ref{sec:FM_fom}.
\item
{\bf output.fom} - This vector contains either a single entry ($1-\sigma$ error), in the case where a one-dimensional likelihood function a parameter $\theta_{\mathrm{A}}$ (for example) is being considered or an array of different FoMs when an ellipse of two parameters is being plotted. These are each explained in above in \ref{sec:FM_fom}.
\end{itemize}

\subsection{Generating plots}
\begin{figure}[htbp!]
\begin{flushleft}
\includegraphics[width=8cm]{\fig_dir{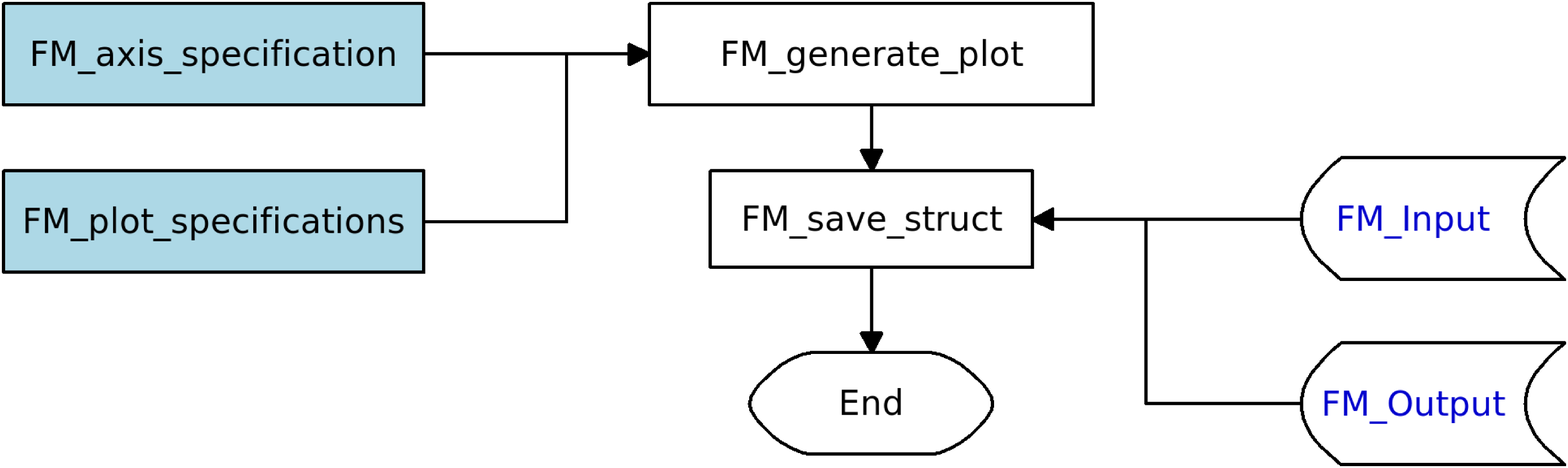}}
\end{flushleft}
\end{figure}
{\bf FM\_generate\_plot} calls the either {\bf FM\_plot\_ellipse.m} or {\bf FM\_plot\_likelihood.m} depending on whether a $1-$D likelihood or an ellipse is required (whether one or two parameters are specified in the {\bf parameters\_of\_interest} field in the {\bf input} structure. The style of the plot is controlled by the {\bf FM\_plot\_specifications.m} file, which controls variables such as the line style, the colour of the lines, the resolution of the grid and the contour level (for example $1-\sigma, 2-\sigma$). Similarly the file {\bf FM\_axis\_specifications.m} controls the $x$ and $y$ labels and the range of the plot that will be generated.

Lastly {\bf FM\_save\_struct} is called to save the {\bf input} and {\bf output} structures with a user specified filename. One could invoke this function from the command line:
\begin{verbatim}
 >>FM_save_struct(`saved_filename',input,output)
\end{verbatim}
where {\bf input} and {\bf output} correspond to the structures that are being saved as {\bf saved\_filename-01-Nov-2008.mat}. The date on which the structure is saved is appended to the end of the filename. If a filename is not specified then the default name of {\bf FM\_saved\_data} is used. It is important to note that the function overwrites existing files with the same name and date and no warning is given. Care should thus be taken to ensure different names are specified when saving important data on the same day. The structures can be loaded once again by issuing the following command:
\begin{verbatim}
 >>load(`saved_filename-01-Nov-2008')
\end{verbatim}
this will make the previously used {\bf input} and {\bf output} structures available in the current session.

\begin{figure}[htbp!]
\begin{center}
\includegraphics[width = 0.45\textwidth]{\fig_dir{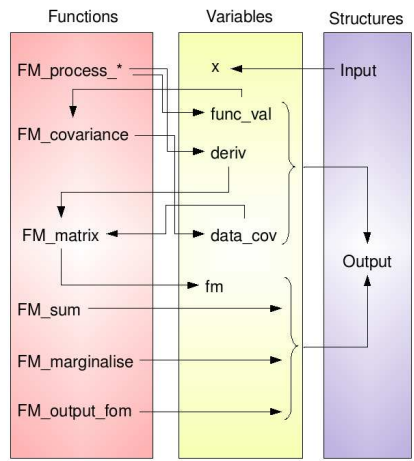}}
\caption{Schematic of the relationships between the various functions, variables and structures of the code. The vertical bars indicate the functions, variables and structures respectively while the arrows illustrate how a function may produce a variable and how that variable is in turn stored in a structure (left to right). Alternately the arrows can also show how a variable is retrieved from a structure and (possibly) used in a function. The order from top to bottom shows the chronological order in which the functions, variables and structures are called {\bf FM\_run.m}.}
\label{links}
\end{center}
\end{figure}

To end with we provide a global view of the structure of the code in Figure~(\ref{links}).

\section{Cosmological Application}
\subsection{The Background Observables \label{obs}}

{\em -- Hubble parameter}\\
The expansion history of the Universe is described by the Hubble parameter, which is defined as\begin{equation}
H^2(z) = H_0^2 E^2(z) = H_0^2\left( \om (1+z)^{3} + (1-\om-\ok) f(z, w_0, w_a) + \ok(1+z)^{2}\right),
\label{eeq}
\end{equation}

with the evolution of the dark energy density, $f(z)$ given by
\begin{equation}f(z) =\exp \left(3 \int_0^z \frac{1+w(z')}{1+z'}dz' \right). \end{equation}
Assuming the Chevallier-Polarski-Linder (CPL) expansion of the dark energy equation of state \cite{cp, linder_w}:
\begin{equation}
w(z) = w_0 + w_a\frac{z}{1+z} = w_0 + w_a\left(1-\frac{a}{a_0}\right)\label{wparam}\,,
\end{equation}
where $a_0 = c/(H_0\sqrt{|\Omega_k|})$ is the curvature radius of the cosmos, $f(z)$ becomes:
\begin{equation}
f(z)= (1+z)^{3(1+w_0+w_a)} \exp \left\{-3w_a \frac{z}{1+z}\right\}\label{feq}\end{equation}

{\em -- Angular diameter distance}\\
Measurements of `standard rulers' of known intrinsic length is widely used as a probe of the cosmology of the Universe. The angular diameter distance relates the angular size of an object to its known length to obtain a measure of the distance to the object, and given by

\begin{equation}
d_A(z) = \frac{1}{1+z} \frac{c}{H_0} \frac{1}{\sqrt{\ok}}.\sinh \( \sqrt{\ok} \chi(z) \),
\label{daeq}
\end{equation}

where \begin{equation}
\chi(z) = \int_0^z \frac{1}{E(z')} dz',
\label{chiz}
\end{equation}
and $E(z)$ is as defined in Eq.~(\ref{eeq}).\\

{\em -- The Growth of Structure}\\
The growth of structure is a potentially powerful probe of dark energy \cite{correct_detf, amendola, amendola07, loeb_growth, wang_growth,linder_growth05, linder_growth09}. Consider the differential equation for the evolution of perturbations in the matter density $\delta$ (assuming the pressure and pressure perturbations of the matter are zero - $p=\delta p=0$) \cite{peebles1993,wang_steinhardt,linder_jenkins}:
\begin{equation}
\ddot{\delta} + 2H\dot{\delta} = 4\pi G\rho_m \delta
\label{deltaeq}
\end{equation}

The growth function provides the temporal evolution of these density perturbations, i.e. $\delta({\bf x},z) \propto G(z)$. In \name~his is solved in a general FLRW universe, and hence there is in general no analytical solution to Eq.~(\ref{deltaeq}). Under the assumption of a {\em flat universe} and a cosmological constant (or pure curvature) however, the growing mode satisfies the following integral form \cite{growth_eis, heath77}:
\begin{equation}
G(z) =\frac{5 \om E(z)}{2} \intzinf \frac{(1+z')dz'}{E(z')^3},
\label{growthz}
\end{equation}
where the $5/2$ coefficient is chosen to ensure that $G(z) \rightarrow 1/(1+z)$ as $z \rightarrow \infty$
This expression should not be used however, to compute the Fisher derivatives $\partial G/\partial \Omega_k, \partial G/\partial w_0$ or $\partial G/\partial w_a$ since all of these derivatives violate the validity of the equation. Instead, the growth derivatives should be computed numerically from the solution of the full differential equation for $\delta(x)$. Rewriting the Raychaudhuri equation in terms of the Friedmann equation and the curvature density allows one to find an equation explicitly showing the curvature and dynamical dark energy contributions to the friction term: \be
G'' + \frac{3}{2}\left(1 +\frac{\Omega_k(x)}{3} - w(x)\Omega_{\mathrm{DE}}(x)\right)\frac{G}{x} -\frac{3}{2}\left(\Omega_m(x)\right)\frac{G}{x^2} = 0, \label{newg}
\ee
where the new independent variable is $x \equiv a/a_0 = 1/(1+z)$, $a_0$ is the radius of curvature and $\Omega_k(x) = -k/(a_0^2x^2H(x)^2)$ \& $\Omega_{\mathrm{DE}}(x) = \rho_{\mathrm{DE}}(x)/\rho_{crit}(x)$ are the fractions of the critical density in curvature and dark energy respectively. Alternatively, this can be written as a differential equation in terms of $\ln(x)$:
\be
\frac{d^2 G}{d\ln^2 x} + \frac{3}{2}\left(\frac{1}{3} +\frac{\Omega_k(x)}{2} - w(x)\Omega_{\mathrm{DE}}(x)\right)\frac{d G}{d\ln x} -\frac{3}{2}\Omega_m(x)G = 0, \label{newgln}
\ee
which is the equation actually solved in \name~since it is typically more stable numerically. Appropriate initial conditions for this differential equation are set deep in the matter dominated era $G(z_i) = 1, dG/d\ln x(z_i) = G(z_i)$ for $z_i \geq 100.$\footnote{In particular one can compare the results with the growth code available at\\http://gyudon.as.utexas.edu/$\sim$komatsu/CRL/.} Note that as a result, the growth solutions will be
unreliable if $w(z \rightarrow \infty) = w_0 + w_a \geq 0$ (or even if it is close to zero from below) since then there will be significant or even dominant early dark energy. \name~allows the user to choose the redshift where the growth is normalised to unity. The Fisher derivatives all satisfy $\partial G/\partial \theta_i = 0$ at the normalisation redshift.

\subsection{Alternative Dark Energy Parametrisations}
The \name~GUI is hard-coded for three cosmological observables ($H$,$d_A$, and $G$), assuming the Chevallier-Polarski-Linder (CPL) parameterisation \cite{cp,linder_w} with parameters ($w_0,w_a$) -- see Eq.~(\ref{wparam}). This is true of both the functions themselves, and the analytical derivatives included in the \name~suite. The general framework of \name, however, means that one is not restricted to this parametrisation. As can be seen from Eq.~(\ref{eeq}), (\ref{daeq}) and (\ref{newgln}), the dark energy equation of state enters in the cosmological observables through the evolution of the dark energy, via $f(z)$, defined in Eq.~(\ref{feq}). Hence for any given $w(z)$ one only needs to specify (in the {\bf input} structure) the names of the functions that will replace the current versions of {\bf FM\_function\_1.m} (H), {\bf FM\_function\_2.m} (DA) and {\bf FM\_function\_3.m} (G). The same is true for the derivatives - either they can be coded analytically for the particular parametrisation of dark energy, or the derivatives will be evaluated numerically from the functions specified in the {\bf input} structure.

As a caveat, the GUI can only be used if the new parametrisation of dark energy still contains only two coefficients. If this is not the case, \name~must be run from the command line version.

\section{Extensions}
The general philosophy of \name{}~was to make it as easy as possible to mould and extend to the needs of a general user. In line with this philosophy we have introduced extensions as a means to add functionality and customisation to the existing \name{}~suite. As a design philosophy for future extensions we envisage that extensions do not alter the core functions of the code but rather access the input, output or other modular core-functions. This will enable a large community of contributors to add and make available their own specific extensions while ensuring that the robust design features of the core code remain intact.

An important element in ensuring the success of shared extensions is that all contributors have a good appreciation of the structures used in \name{}~while also documenting and commenting the details of their code thoroughly, including the purpose of the extension, the required input and output produced, which files or structures the extension interacts with from \name{}~and whether it is run from the command line or GUI.

The latest extensions included in this release are listed below.

\subsection{Obtaining Baryon Acoustic Oscillation Errors from Survey Parameters}
Two modules are included that calculate errors on the Hubble parameter and angular diameter distance in BAO surveys characterised in input structures of survey parameters. The provided codes are extensions to \name{}, and should be placed in the same folder as the main code suite so that they can access the required elements of the \name{}~suite.

The first of these extensions, {\bf EXT\_FF\_Blake\_etal2005} uses the fitting formulae of Blake {\em et al.} \cite{blake_etal_ff} to calculate the errors on the Hubble parameter and angular diameter distance given certain survey specifications, such as the survey area and the redshifts used for the measurements of $H$ and $d_A$. These are either given as central redshift bins or as the edges of redshift bins. The galaxy number density is also required, and is expected in units of $10^{-3}\mathrm{Mpc}^{-3}h^3$.

All files associated with this module have the same prefix to identify them as external modules for the fitting formula of Blake {\em et. al} \cite{blake_etal_ff}. The fitting formulae contain coefficients specific to either photometric or spectroscopic surveys, and hence it must be specified which survey one is considering. The input parameters to this module are supplied in an input structure, which is explained in the {\bf EXT\_FF\_Blake\_etal2005\_Readme.txt} file. They are:
\begin{itemize}
 \item Input\_survey.base\_parameters: The fiducial values of the cosmological parameters are specified here, as ($H_0, \Omega_m, \Omega_K, w_0,w_a$), in the same way as the input parameters are defined in the main \name{}~code.
\item Input\_survey.surv\_type: This specifies either a photometric or spectroscopic survey by either setting it to  `spec' or `phot'. This defines which of the sets of coefficients to use in the fitting formulae.
\item Input\_survey.z\_type: This field indicates that the redshift data being used are either given at the `edge' of the bins or as the `central' redshifts. If Input\_survey.z\_type=`central' then an additional input for width of the redshift bins, dz, is required.
\item Input\_survey.area: This field specifies the area of the survey in units of 1000 square degrees.
 \item Input\_survey.n: This gives the number density of galaxies and is measured in units of $10^{-3}$h$^3$Mpc$^{-3}$
\item Input\_survey.vecH and Input\_survey.vecDA: These entries give the redshift vectors of the survey for the Hubble parameter $H(z)$ and angular diameter distance. $d_A(z)$
\item Input\_survey.biasH and Input\_survey.biasDA : These entries specify the bias for the Hubble parameter and angular diameter distance. These two vectors should have the same length as that of `Input\_survey.vecH' and `Input\_survey.vecDA' respectively.
\end{itemize}

\begin{figure}
\centering
 \includegraphics[width = 4in]{\fig_dir{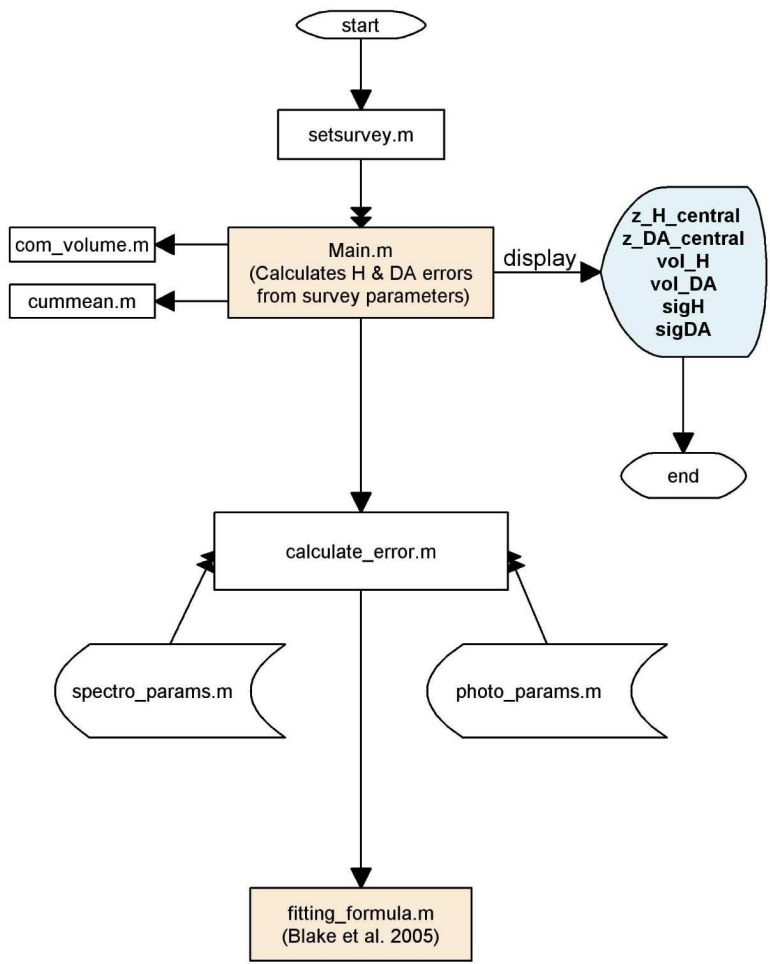}}
\caption{The flow of inputs and functions within the Blake {\em et al.} \cite{blake_etal_ff} extension module to \name{}.\label{fig:ext_flow_blake_et_al}}
\end{figure}

An example of a default input structure supplied in the file {\bf EXT\_FF\_Blake\_etal2005\_Input.m}. The command used to run the Blake {\em et al.} \cite{blake_etal_ff} fitting formula is given as:
\begin{verbatim}
 >>[z_H_central,z_DA_central,vol_H, vol_DA,sigH, sigDA] = ...
 >>EXT_FF_Blake_etal2005_Main(Input_survey)
\end{verbatim}
Should no input structure be specified when calling the code, the default input ({\bf EXT\_FF\_Blake\_etal2005\_Input.m}) is assumed. The module comprises of various smaller modules to compute the oscillation scales in the radial and transverse direction, and hence the errors on the angular diameter distance and Hubble parameter. Figure~(\ref{fig:ext_flow_blake_et_al}) outlines the procedures in the module. The module outputs the following parameters:
\begin{itemize}
 \item z\_H\_central: A vector of the central redshifts of the $H(z)$ data bins, related to the Input\_survey.vecH fields specified in the input structure.
\item z\_DA\_central: A vector of the central redshifts of the $d_A(z)$ data bins, related to the Input\_survey.vecH fields specified in the input structure.
\item vol\_H: A vector of the volume of the redshift bins for the Hubble parameter.
\item vol\_DA: A vector of the volume of the redshift bins for the angular diameter distance.
 \item sigH: A vector of the {\em fractional} error on the Hubble parameter: $\sigma_H/H$. For percentage errors, multiply this by 100.
 \item sigH: A vector of the {\em fractional} error on the angular diameter distance: $\sigma_{d_A}/d_A$. For percentage errors, multiply this by 100.
\end{itemize}

The module {\bf EXT\_FF\_SeoEisenstein2007} also computes the errors on the Hubble parameter and angular diameter distance using the prescription set out in \cite{SE2007} and the sound horizon scale as given in \cite{wangsh}. This module contains a wrapper to call the Matlab version of the {\bf C} code of Seo \& Eisenstein \cite{SE2007code}. In this module the code does not need to be in the same directory as the \name{}~suite, and runs completely independently of \name{}.

The module also takes an input structure (a default input structure with all fields specified is given in {{\bf EXT\_FF\_SeoEisenstein2007\_Input.m}}) with the following parameters defined:
\begin{itemize}
 \item {\bf Input\_survey.number\_density}: The galaxy number density in units of h$^3$ Mpc$^{-3}$
\item {\bf nput\_survey.wmap}: A flag (1 or 3) specifying which WMAP power spectrum to use.
\item {\bf Input\_survey.sigma8}: The value of $\sigma_8,$ the amplitude of linear clustering on a scale of 8 Mpc.
\item {\bf Input\_survey.Sigma\_z} The line of sight real mean squared comoving distance error due to redshift uncertainties.
\item {\bf Input\_survey.beta}: The redshift distortion parameter is entered.
\item {\bf Input\_survey.volume}: The survey volume in units of h$^{-3}$ Gpc$^3$.
\end{itemize}

Sigma\_perp (the transverse rms Lagrangian displacement) and Sigma\_par (the radial displacement) are calculated and saved to the input structure. The module is comprised of smaller functions; the flowchart of the module is shown in Figure~(\ref{fig:ext_flow_seo_eisenstein}).

As in the case of the Blake {\em et al.} module, the Seo and Eisenstein module code can be called from the command line using:
\begin{verbatim}
 >>[Drms,Hrms,r,Rrms] = EXT_FF_SeoEisenstein2007_Main(Input_survey)
\end{verbatim}
\begin{figure}
\centering
 \includegraphics[width=4.2in]{\fig_dir{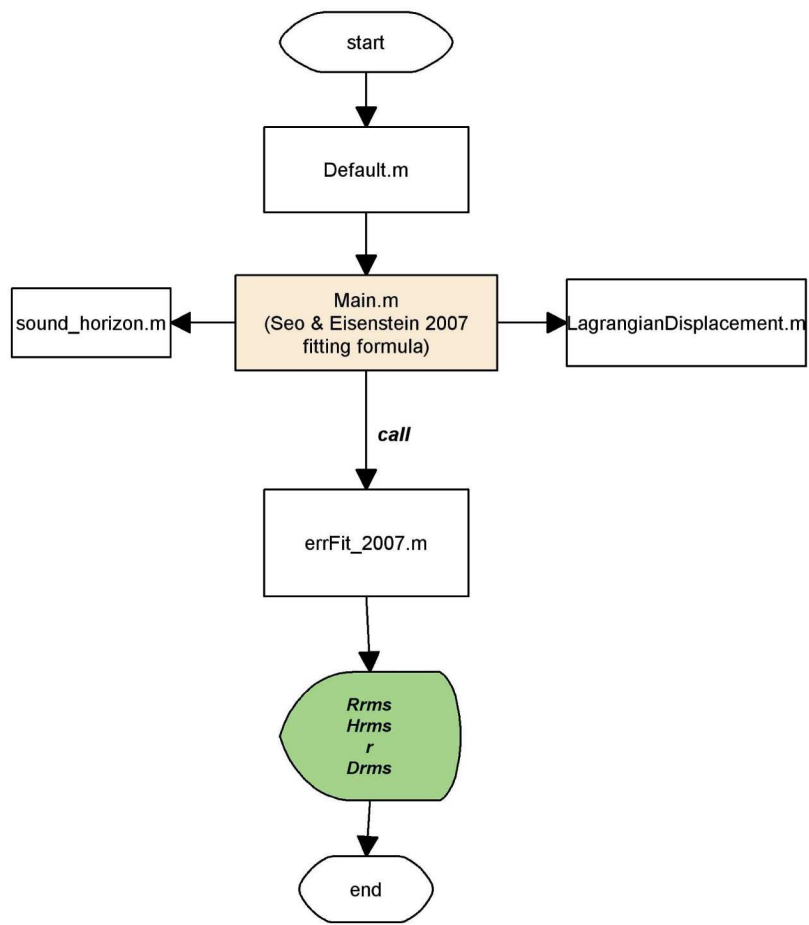}}
\caption{Flowchart of the code in the Seo and Eisenstein BAO extension module to \name.\label{fig:ext_flow_seo_eisenstein}}
\end{figure}
The outputs of the code are the root mean square error on $D/s$ and $Hs,$ where $s$ is the oscillation scale. These are both given as fractional error, for percentage error multiplied by 100. In addition the correlation coefficient between $D$ and $H$ is given ($r$), and the diagonal entry in the covariance matrix between $D$ and $H$.

To be sure that the extensions have access to the functions contained in \name{}~we need to be sure that the extensions are either placed directly in the same folder or the path is specified to both \name{}~and the extensions. One can use the path command to do this:
\begin{verbatim}
 >>path(path,`/path-to-folder/Fisher4Cast-v2.0')
 >>path(path,`/path-to-folder/EXT_FF_Blake_etal2005')
\end{verbatim}
where the `path-to-folder' is the path specifying the directory where the extension or \name{}~code is kept on your local computer.
Similarly, to run the Seo and Eisenstein  \cite{SE2007} module, you will need to ensure that the respective extension is either in the same directory or the path is specified:
 \begin{verbatim}
 >>path(path,`/path-to-folder/Fisher4Cast-v2.0')
 >>path(path,`/path-to-folder/EXT_FF_SeoEisenstein2007')
\end{verbatim}

\subsection{Reporting Features for the \name{}~Suite\label{report}}
Two extension modules have been included to provide reports of the \textbf{input} and \textbf{output} structures during a run of \name{}. These reports can either generate an ASCII text file (.txt) or a \LaTeX{}~file (.tex) which detail all the input and output produced by \name{}.

In the case of the \LaTeX{}~reporting function the resulting .tex file can be compiled using \LaTeX{}~to produce a Postscript file (.ps) or Portable Document Format file (.pdf). This allows for a more polished presentation of the results generated from \name{}. It also includes a figure of the ellipse or likelihood plot which is embedded in the document. The additional benefit to generating a document in .tex format is that one can cut-and-paste the \LaTeX{}~formatted syntax of the figure or any of the tabulated data for easy inclusion of the figure in an article or document containing the results from a run of \name{}.
\begin{figure}[h!]
\centering
\begin{tabular*}{\textwidth}{c r}
 \includegraphics[width=0.45\textwidth]{\fig_dir{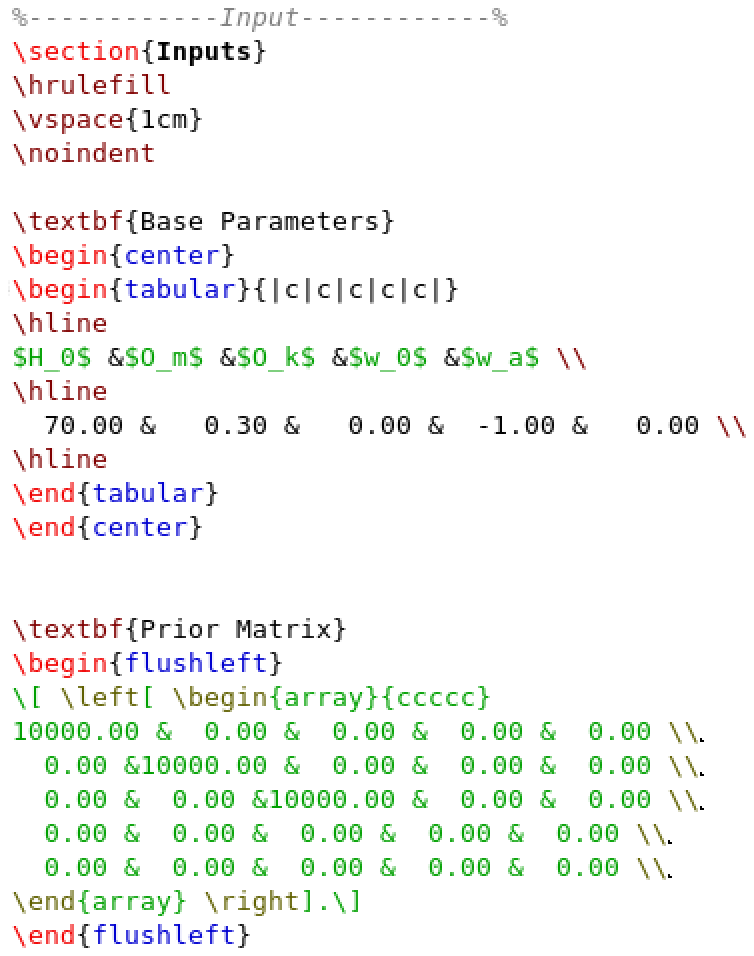}}&
 \includegraphics[width=0.45\textwidth]{\fig_dir{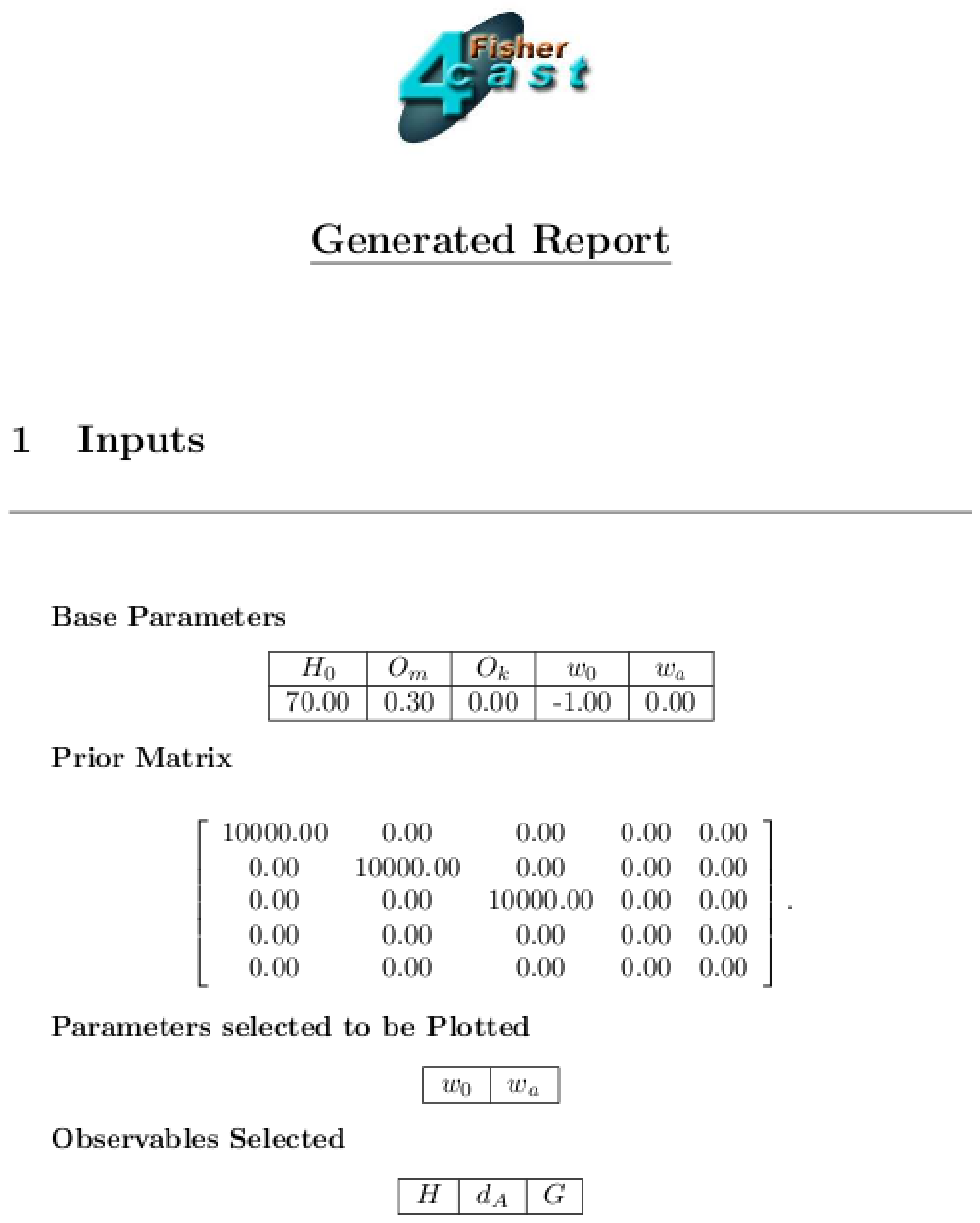}}\\
 \includegraphics[width=0.45\textwidth]{\fig_dir{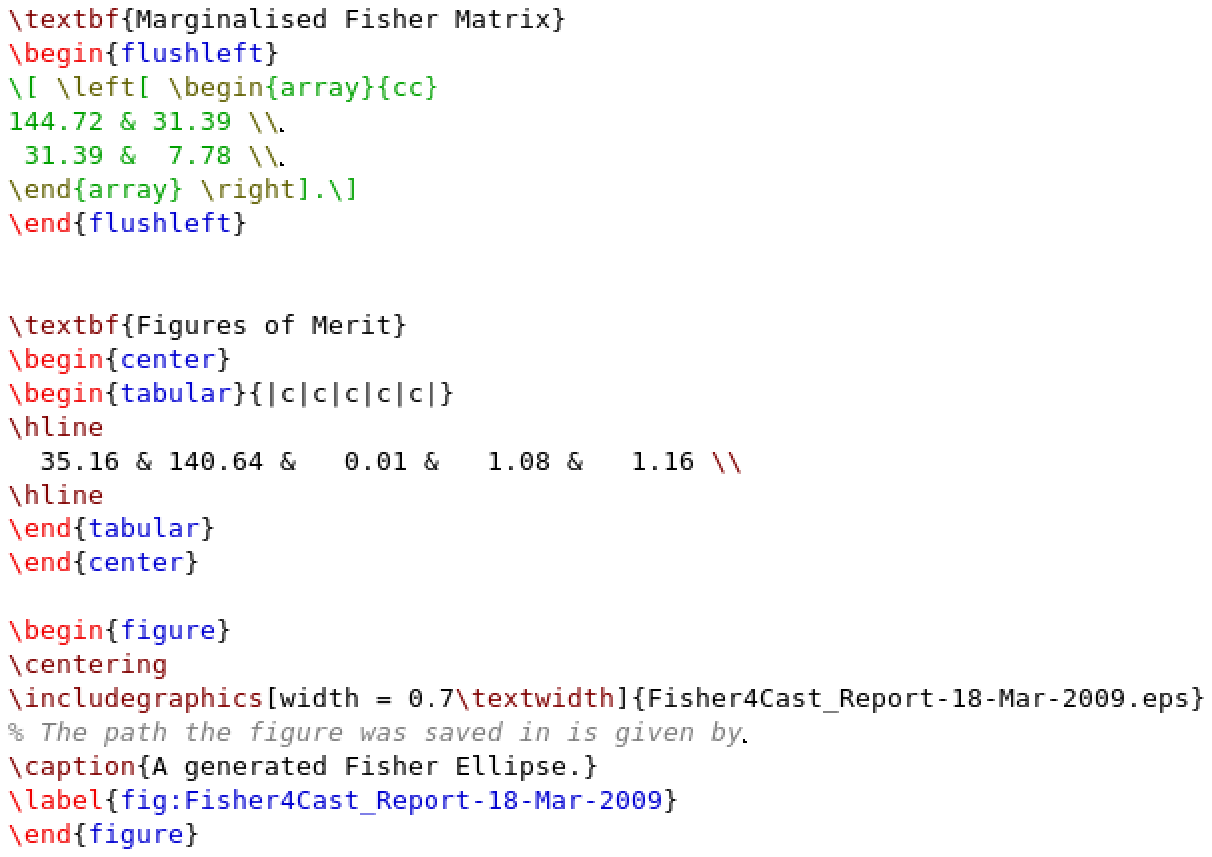}}&
 \includegraphics[width=0.45\textwidth]{\fig_dir{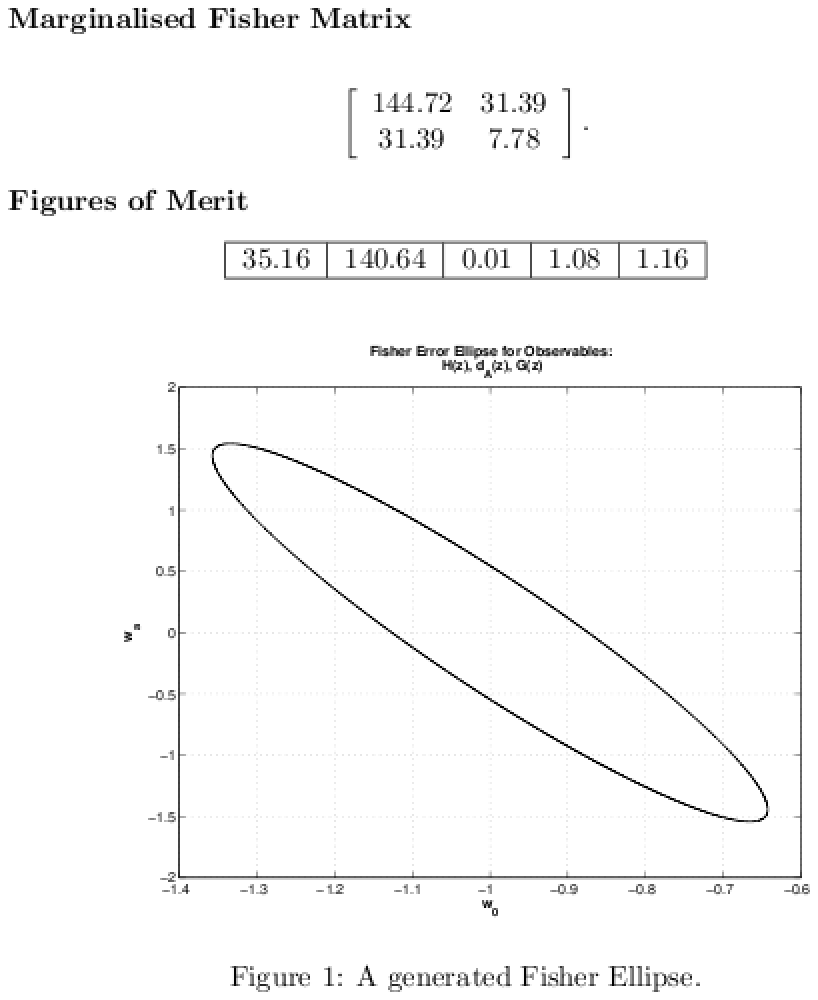}}
\end{tabular*}
\caption{The figures on the left show excerpts from the .tex files that was generated from the reporting code while the figures on the right show the extracts of the corresponding .pdf file that was generated from the .tex file. \label{fig:report_latex}}
\end{figure}
These reporting features are accessible through the Graphical User Interface, by clicking the scroll-down menu bar labelled `Saving Features' (see Figure~(\ref{11})) for a screenshot of the drop-down menu). This opens a dialog box which in the case of the text report prompts the user for the .txt filename that it should be saved as. Upon choosing a \LaTeX{}~report, two dialogue boxes are opened and the user is prompted for the names of both the .tex file and the .eps file.

These two extensions can just as easily be called from the command line. To generate a text report one uses the functions \textbf{FM\_report\_text.m}. The user is required to supply at least an input structure to generate a report. This input structure can either be a default input structures, eg \textbf{ Cooray\_et\_al\_2004.m}, or a user customised input. The function \textbf{FM\_report\_text.m} then calls {\bf FM\_run(input)} with the same supplied input which then produces the relevant output structure. Both the input and output used and generated from \name{}~are then recorded in the report. The user can also specify a filename to saved the report as (if no .txt extension is supplied one will be added automatically). A default name of 'Fisher4Cast\_Report-Day-Month-Year.txt' will be used, should no name for the report be specified, where the Day-Month-Year are the date on which the report was generated. For example the command:
\begin{verbatim}
>>FM_report_text(input,`report_name')
\end{verbatim}
will generate a report with the name, `report\_name.txt', as described above. If the same report\_name is used, the previous report will be  overwritten without warning. Please specify a unique report\_name to ensure the report is correctly saved.

Finally there is an option of including a specific output structure in the report function. This is useful when generating the report from the GUI, but care should be taken when using this option in the command line, as one runs the risk of generating a report where the input and output are not appropriately related.
In other words, \begin{verbatim}
>>FM_report_text(input,`report_name',output)
\end{verbatim}
generates a report as before with the name, `report\_name.txt', using the input supplied and {\em assuming} that the given output is associated with the respective input.

Much the same as the text report, the \LaTeX{}~report is called using \textbf{FM\_report\_latex.m} and requires at least an input structure. The filename the report is to be saved as can also be specified (either with or without the .tex extension). The \LaTeX{}~report includes the EPS figure generated from \name{}: as a default the figure will be saved with the same name as the .tex, except an .eps extension. The default name of 'Fisher4Cast\_Report-Day-Month-Year.tex' is used, should no report name be specified The commands
\begin{verbatim}
>>FM_report_latex(input,`report_name')
\end{verbatim}
generate a report with the name, `report\_name.tex', and a figure with the name `report\_name.eps', where the names overwrite any existing files of the same name. Additionally, one can use a specific figure in the report with the command:
\begin{verbatim}
>>FM_report_latex(input,`report_name',`use_fig')
\end{verbatim}
In this case there is of course no guarantee that the figure and the output from \name~agree.

As in the case of the .txt report, one can specify the output structure directly with:
\begin{verbatim}
>>FM_report_latex(input,`report_name',`use_figure',output)
\end{verbatim}
which generates a report as before with the name, `report\_name.tex', using a figure called `use\_figure.eps' where the output and figure are assumed to be associated with the respective input supplied.

\section{How to produce small, good quality postscript images for
inclusion in \LaTeX~documents}

Much of the output from \name{} is expected to be used in research publications produced using \LaTeX, in which case \name{} figures need to be saved in `.eps' or `.ps' format. Unfortunately the default Matlab `.eps' and `.ps' files produced tend to be large, often several MB in size\footnote{A brief word of caution: if you
plot many ellipses in the GUI, or with multiple colour variations, and then save the result, Matlab will save the entire history, making the resulting file large. It is best to decide on what combinations you want and then save only that figure.}.
Apart from making printing slow, this is a problem when submitting papers to online archives,
like the arXiv\footnote{http://xxx.lanl.gov}, which has a strict file limit.

Many of the figures in this paper exceeded 1MB after saving from Matlab, and several exceeded 2MB.
To solve this, one can instead save the files as bitmaps and then use a utility such as {\tt jpeg2ps} to
convert the `.jpeg' file to postscript with a much smaller file size\footnote{This process is discussed at
http://aps.arxiv.org/help/bitmap/index and elsewhere.}. Nevertheless, achieving good compression and good quality results can be tricky to achieve with Matlab figures, and we share the steps that have worked
well, here.

\begin{itemize}
\item Save the files in `.eps' format in Matlab.
\item Load the eps file into Photoshop (or equivalent such as GIMP), which will immediately request to rasterize the file and hence request an image size (in cm or inches) and a resolution (in dots per  inch, dpi).
\item Choose the size to be that which you want the figure to appear in your \LaTeX document (e.g. 10cm). As with all bitmaps, choose it to be the physical size you will use in the end since rescaling bitmaps causes blurring and poor quality. 300dpi gives good resolution. The resulting file will be huge in Photoshop, but do not worry since it is only an intermediate step.
\item Now save the file as a `.jpeg' with high quality factor, $80\%$ or better.
\item Using {\tt jpeg2ps} or a similar utility, convert the `.jpeg' file to `.eps'. This will add approximately $10\%$ to the jpeg file size due to the postscript wrapper. You should be able to achieve a 10-20 fold reduction in eps to eps file size with only marginal reduction in quality.
\end{itemize}

\section{Tests of the Code \label{testpart}}
Various tests were performed to check the correctness and accuracy of all components of \name{}, namely the integration routines in \name, the derivatives and their validity (especially relevant for the growth function) and the matrix manipulation and generation of Fisher error ellipses.

\subsection{Integration Tests}
The integration routines in \name~are tested by comparing them to standard results and fitting formulae for the angular diameter and growth function respectively.

\subsection{Angular Diameter Distance}
\begin{figure*}[htbp!]
\begin{center}
\includegraphics[width = 0.5\textwidth]{\fig_dir{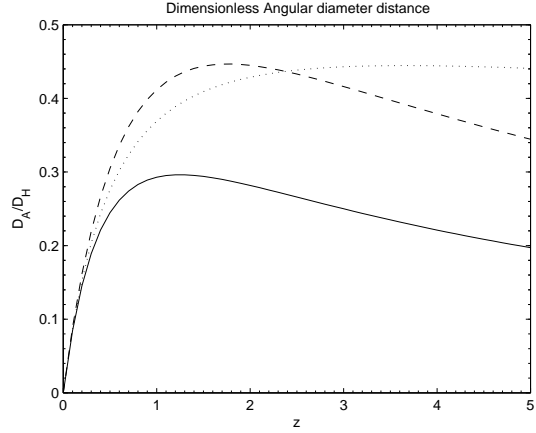}}
\caption{{\bf The angular diameter distance -} plotted for Universes with $(\Omega_m, \Omega_\Lambda) = (1, 0)$ {(solid)}, $(0.05, 0)${(dotted)} and $(0.2, 0.8)${(dashed line)}. Our results are in agreement with a similar plot, Figure (2) in \cite{hogg}.\label{angdidis}}
\end{center}
\end{figure*}

The angular diameter distance, $d_{A}(z)$, defined in Eq.~(\ref{daeq}), is the
ratio of an object's physical transverse size to its angular size (in radians). Characteristically it does not increase indefinitely as $z\rightarrow\infty$, rather it inverts at $z\sim 1$ -- thereafter more distant objects actually appear {\em larger} in angular size. In the case of $\Omega_{\Lambda}=0$, one can write Eq.~(\ref{daeq}) as an analytical function of redshift and cosmic parameters as \cite{weinberg1972,peebles1993}:
\begin{equation}
d_A(z)=\frac{c}{H_0} \,\frac{2\,[2-\Omega_m\,(1-z)-(2-\Omega_m)\,\sqrt{1+\Omega_m\,z}]}{\Omega_m^2\,(1+z)}.
\end{equation}
The angular diameter distances for three particular cosmologies (the same cosmologies as found as Figure (2) of \cite{hogg}) are shown in Figure~(\ref{angdidis}). The same axis lengths and line styles are used for easy comparison of the two plots, which show agreement between the \name~algorithms and the know results.

\subsection{Growth function}
\begin{figure}[htbp!]
\begin{center}
$\begin{array}{@{\hspace{-0.25in}}c@{\hspace{-0.3in}}c}
\includegraphics[width = 3.5in]{\fig_dir{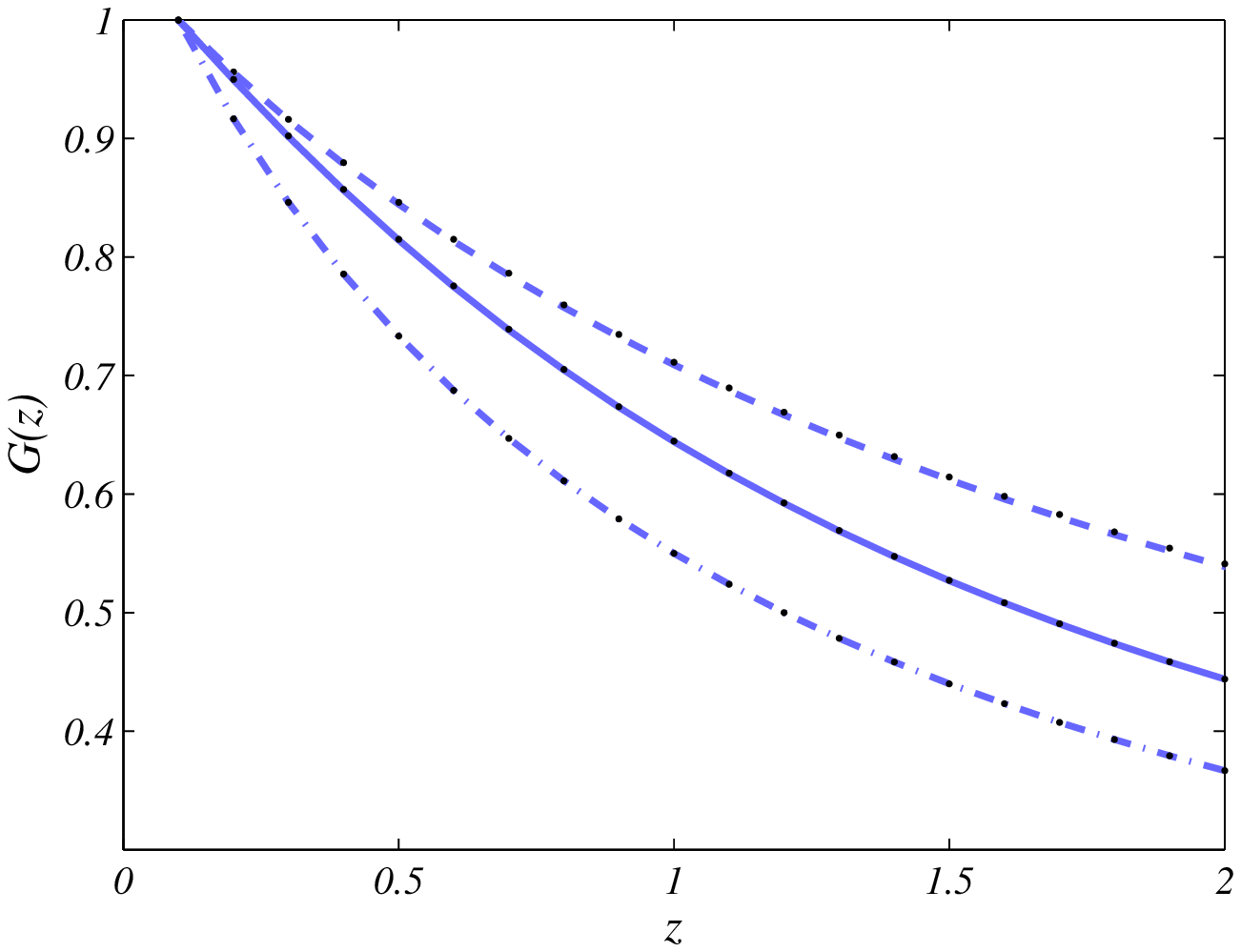}} &
\includegraphics[width = 3.5in, height = 2.5in]{\fig_dir{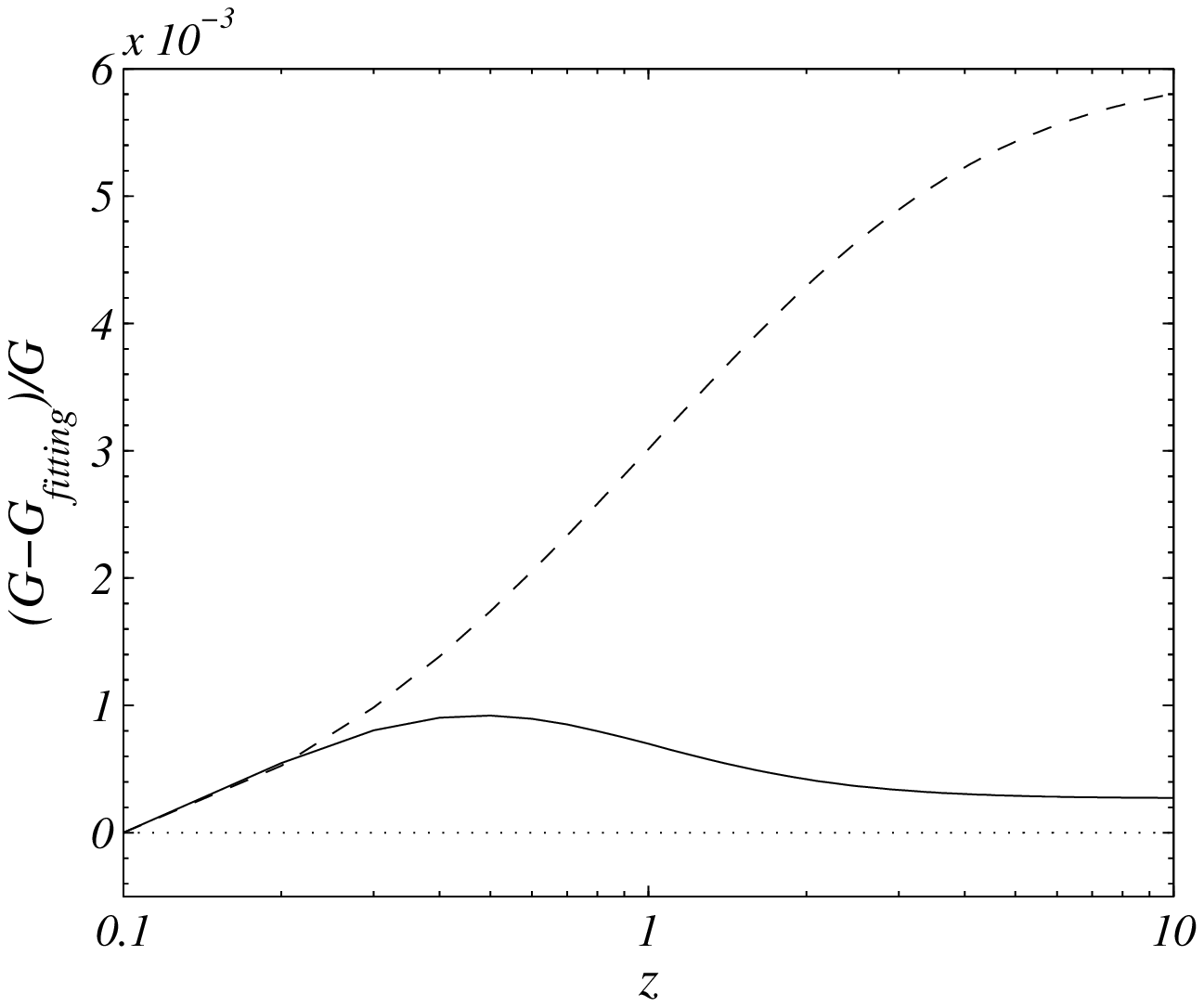}} \\ [0.0cm]
 \end{array}$
\caption{{\bf The \name~growth function compared to a fitting formula.} The left-hand panel shows the growth function implemented in \name~{(blue curves)} as compared to the fitting formula Eq.~\eqref{eq:gFitting} \cite{peacock_dodds,lahav_suto} {(black dots)}. The models represented are $(\Omega_m, \Omega_k) = (0.3, 0.7)$ {(dashed)}, $(0.3, 0)$ {(solid)} and $(1, 0)$ {(dot dashed)}. The normalised differences between the growth function used in \name~and the fitting formula for the various models are of order $10^{-3},$  which is shown in the residuals of the right-hand panel. \label{fig:growth}}
\end{center}
\end{figure}

In order to test the numerical growth function, the solution is compared to other numerical solutions in the literature (code is available for comparison at http://gyudon.as.utexas.edu/\textasciitilde komatsu/CRL/, which implements the growth equation given as Eq.~(76) in \cite{wmap5}), to analytical approximations for particular cosmologies, such as Eq.~(\ref{growthz}), or to fitting formulae, such as that originally suggested by Carroll, Press \& Turner \cite{Carroll_1992}. This fitting formula is given by $G(z) = \frac{g(z)}{1+z}$ \cite{lahav_suto}
where
\begin{equation}\label{eq:gFitting}
g(z) = 5\frac{\Omega_m(z)}{2} \( \Omega_m^{4/7}(z) - \gamma(z) +
\left[1+\frac{\Omega_m(z)}{2}\right] \left[{1+\frac{\gamma(z)}{70}}\right] \)^{-1}
\end{equation}and \begin{eqnarray*}
\gamma(z) &=& \ode\left[\frac{H_0}{H(z)} \right]^2 \\
&=&\frac{\ode}{\om(1+z)^3 + (1-\om - \ode)(1+z)^2 + \ode}.\end{eqnarray*}
The comparison between the growth function from \name{}~and the fitting formula is shown in the left-hand panel of Figure~\ref{fig:growth}, together with the relative difference, $(G - G_{fit})/G,$ between the two methods, in the right-hand panel.

\subsubsection{Degeneracy Tests}
The degeneracy direction of a Fisher ellipse (we consider $w_0, w_a$) is a useful diagnostic of whether or not the Fisher ellipses are being calculated correctly. The direction of degeneracy can be computed analytically for a given redshift by assuming that the specific observable $X^\alpha$ is constant at a particular redshift, and to then solve for $w_a$ as a function of $w_0$, or by computing the likelihood over a grid of $w_0-w_a,$ and then taking contours of the likelihood that correspond to the same cosmology as that assumed in the run of~\name{}. This is particularly important when considering numerical derivative routines, such as those needed for the growth function.

\subsubsection{Growth function}
\begin{figure}[t]
\begin{center}
$\begin{array}{l@{\hspace{-0.1in}}c@{\hspace{-0.1in}}c}
	\epsfxsize=3.7in
	\epsffile{\fig_dir{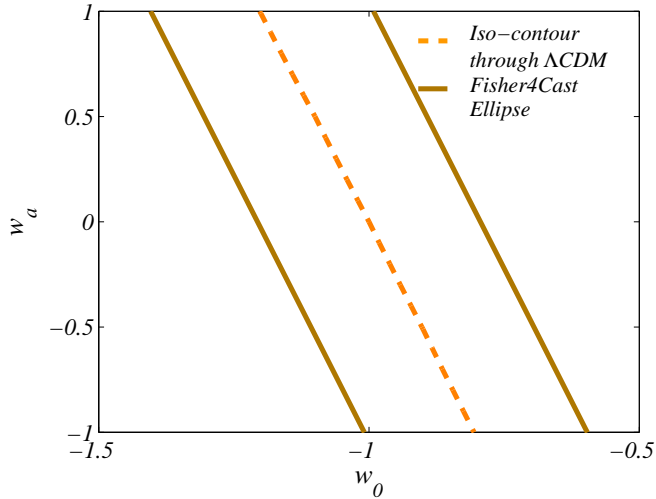}}
 \end{array}$
  \caption{{\bf Grid test of the numerical growth derivatives evaluated at a single redshift:} The growth function is evaluated at a single redshift $z=1$ for a range of models on a grid of $-1.5<w_0<0.5$ and $-1<w_a<1.$ The $w_0-w_a$ degeneracy direction in an assumed $\Lambda$CDM cosmology is then the iso-growth contour corresponding to $\Lambda$CDM ($w_0 = -1, w_a = 0$), shown here as the orange dashed line. The \name{}~degeneracy direction (computed assuming an arbitrary value of $1.5\%$ error on growth) is shown as brown solid lines. \label{growthgrid}}
  \end{center}
 \end{figure}

As a test of the correctness of the solution (and Fisher derivatives) of Eq.~(\ref{newgln}), the Fisher ellipse from a survey consisting of a single measurement of the growth function at $z=3$ (as computed with \name) are shown in Figure~(\ref{growthgrid}), overlaid with contours of the likelihood corresponding to the growth value of the fiducial cosmology. The agreement of the degeneracy directions indicates that numerical computation of both the function and its Fisher derivatives is sound.

\subsubsection{Hubble parameter}
\begin{figure}[htbp!]
\begin{center}
\includegraphics[width=.55\textwidth]{\fig_dir{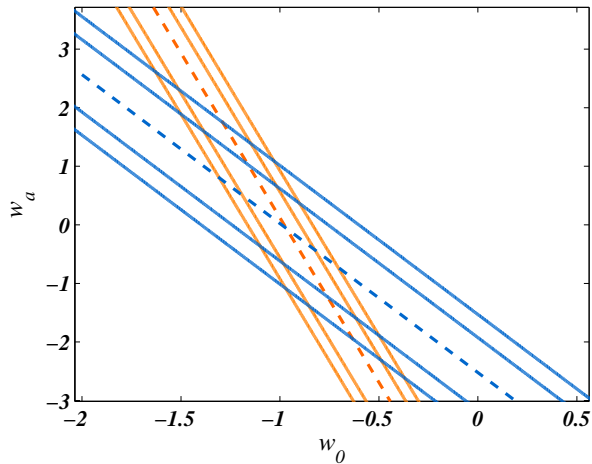}}
\caption{{\bf Lines of degeneracy --} comparing the analytical degeneracy directions for the Hubble parameter {(long dashed lines)} with the degenerate error ellipse produced using \name{}~{(solid lines)} for two separate surveys of $10\%$ measurements of $H$, one at $z = 0.5$, and another at$z= 2$. The analytical solution and that of the error ellipse coincide, which is a systematic check of the computations and correctness of \name. \label{hdegen}}
\end{center}
\end{figure}
In the case of the Hubble parameter one can compute the degeneracy direction analytically: consider a single perfect measurement $H(z)$ at some particular redshift $z.$ Solving for $w_a$ in terms of $w_0$ by substituting Eq.~(\ref{feq}) into (\ref{eeq}), yields
\begin{eqnarray}
C(z) & \equiv & \frac{\ln \left( \frac{H^2}{H_0^2}- \om (1+z)^3 - \ok (1+z)^2 \right)}{\ln(1-\om -\ok)} \nonumber \\
 &=&3(1+w_0 +w_a)\ln(1+z) - 3w_a\frac{z}{1+z},
\end{eqnarray} hence
\be
\label{hwa}
w_a = \frac{C(z) - 3(1+w_0)\ln(1+z)}{3(\ln(1+z) - z/(1+z))}
\ee
This degeneracy direction is shown in Figure~(\ref{hdegen}) for two different values of the central redshift $z$ along with the ellipses from the ellipses from \name{}. The agreement between the analytical solution for the degeneracy direction, and the degenerate ellipse from the \name~code confirms that the derivatives are being calculated correctly, and the matrix operations within \name~are sound.
\subsubsection{Angular Diameter Distance}
\begin{figure}[htbp!]
\begin{center}
\includegraphics[width=3.7in]{\fig_dir{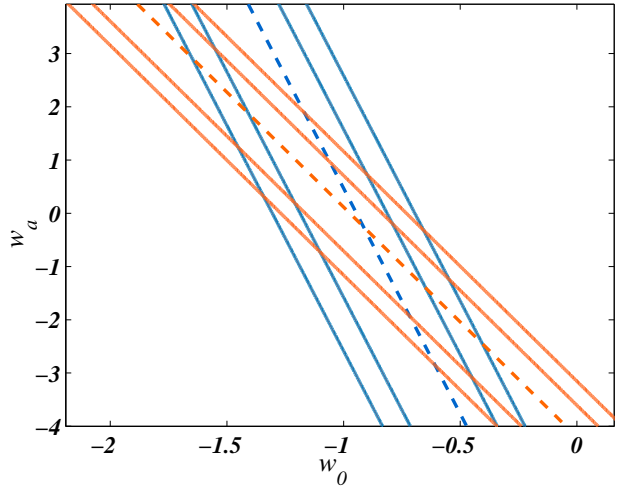}}
\caption{{\bf Lines of degeneracy -- } similarly to Figure~(\ref{hdegen}), the analytical degeneracy directions for the angular diameter distance {(long dashed line)} are superimposed on the degenerate error ellipses produced from \name{}~{(solid lines)} for two surveys consisting of one measurement each of $d_A$ with a $10\%$ error at redshifts $z = 0.5$ and $z= 2$ respectively. \label{dadegen}}
\end{center}
\end{figure}

Furthermore, it is possible to obtain an analytical solution for the direction of degeneracy between the parameters $w_0,w_a$ from measurements of the angular diameter distance, $d_A(z)$ (Eq~(\ref{daeq})). The dark energy parameters only enter Eq~(\ref{daeq}) in $\chi$, the integral of $1/E$, hence we need only take derivatives $ \p \chi_/ \p w_i$ in order to determine the degeneracy, via:
\bea
\label{dawa}
w_a &=& \frac{\p w_a}{\p w_0} w_0 + Q \nonumber \\
&=& \frac{\p \chi/ \p w_0 }{\p\chi /\p w_a } w_0 + Q,
\eea
where $Q$ is an arbitrary constant. Figure~(\ref{dadegen}) once again confirms the agreement between the degeneracy direction of the ellipse produced using \name~and the analytical solution, for two values of the central redshift $z$.

\subsubsection{Combined Degeneracies}
Combining measurements from multiple probes is key to tightening constraints on parameters in a cosmological model. Summing over different observables $X^\alpha$ means that one effectively ``combines'' the different degeneracy directions of each observable together. Given the analytical degeneracy directions discussed earlier, one can investigate (qualitatively at least) how the these degeneracy directions combine to form the Fisher ellipse. This is illustrated for the Hubble parameter and angular diameter distance at a redshift of $z = 0.5$. The orange dashed line shows the degeneracy direction for the angular diameter distance, the degeneracy direction for the Hubble parameter is shown in blue -- the combined ellipse in the $w_0-w_a$ plane for these two observables (black ellipse) is a combination of both degeneracy directions. Qualitatively, the magnitude of the Fisher derivatives for either observable (an indication of how sensitive the observable $X^\alpha$ is to the parameter $\theta_{\mathrm{A}}$) determines how much its degeneracy direction contributes to the final ellipse. This can be used to investigate how the degeneracy directions change with redshift.
\begin{figure}[htbp!]
\begin{center}
\includegraphics[width=3.7in]{\fig_dir{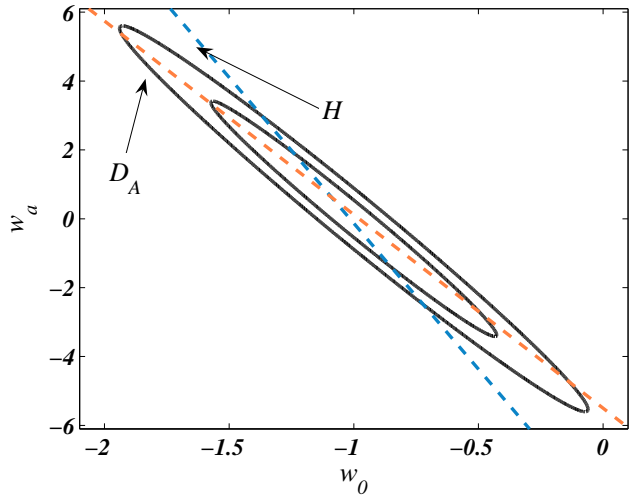}}
\caption{{\bf Contributions to Fisher ellipse} The separate degeneracy directions ({orange dashed line} for $d_A$ and ({blue dashed line} for $H$) are plotted along with the Fisher ellipse for $w_0-w_a$ computed at a redshift of $z=0.5$, illustrating how the various observables contribute to the overall direction of the Fisher ellipse. One can then employ this strategy at varying redshifts to track how the contributions change as a function of redshift, as the observables become more or less sensitive to the dark energy parameters $w_0,w_a$.}
\end{center}
\end{figure}

\bibliographystyle{cup}
\bibliography{artman}

\begin{thebibliography}{35}
\expandafter\ifx\csname natexlab\endcsname\relax\def\natexlab#1{#1}\fi
\expandafter\ifx\csname bibnamefont\endcsname\relax
  \def\bibnamefont#1{#1}\fi
\expandafter\ifx\csname bibfnamefont\endcsname\relax
  \def\bibfnamefont#1{#1}\fi
\expandafter\ifx\csname citenamefont\endcsname\relax
  \def\citenamefont#1{#1}\fi
\expandafter\ifx\csname url\endcsname\relax
  \def\url#1{\texttt{#1}}\fi
\expandafter\ifx\csname urlprefix\endcsname\relax\def\urlprefix{URL }\fi
\providecommand{\bibinfo}[2]{#2}
\providecommand{\eprint}[2][]{\url{#2}}

\bibitem[1]{f4c}
\bibinfo{author}{\bibfnamefont{B.~A.} \bibnamefont{{Bassett}}},
  \bibinfo{author}{\bibfnamefont{Y.~T.} \bibnamefont{{Fantaye}}},
  \bibinfo{author}{\bibfnamefont{R.}~\bibnamefont{{Hlozek}}},
    \bibinfo{author}{\bibfnamefont{J.}~\bibnamefont{{Kotze}}}
 \emph{\bibinfo{title}{Fisher Matrix Preloaded --
  Fisher4Cast, submitted}} (\bibinfo{year}{2009}).

\bibitem[2]{tegmark}
\bibinfo{author}{\bibfnamefont{M.}~\bibnamefont{{Tegmark}}},
  \bibinfo{author}{\bibfnamefont{A.~N.} \bibnamefont{{Taylor}}},
  \bibnamefont{and} \bibinfo{author}{\bibfnamefont{A.~F.}
  \bibnamefont{{Heavens}}}, \bibinfo{journal}{\apj}
  \textbf{\bibinfo{volume}{480}}, \bibinfo{pages}{22} (\bibinfo{year}{1997}) 
  {ArXiv:astro-ph/9603021}.
  
\bibitem[3]{BSD}
\bibinfo{note}{Read the BSD license available in the \name{}~distribution, or
  go to http://www.opensource.org/licenses/bsd-license.php for more
  information.}
  
  

\bibitem[4]{cosmo_org}
\emph{\bibinfo{title}{http://www.cosmology.org.za}}.

\bibitem[5]{mathworks}
\emph{\bibinfo{title}{http://www.mathworks.com/matlabcentral/fileexchange/}}.

\bibitem[6]{millenium_sim}
\emph{\bibinfo{title}{http://www.mpa-garching.mpg.de/galform/millennium/}}.


\bibitem[7]{wmap_back}
\emph{\bibinfo{title}{http://map.gsfc.nasa.gov/}}.

\bibitem[8]{matrix_free}
\emph{\bibinfo{title}{http://commons.wikimedia.org/}}.


\bibitem[9]{cooray2004}
\bibinfo{author}{\bibfnamefont{A.}~\bibnamefont{{Cooray}}},
  \bibinfo{author}{\bibfnamefont{D.}~\bibnamefont{{Huterer}}},
  \bibnamefont{and}
  \bibinfo{author}{\bibfnamefont{D.}~\bibnamefont{{Baumann}}},
  \bibinfo{journal}{\prd} \textbf{\bibinfo{volume}{69}},
  \bibinfo{pages}{027301} (\bibinfo{year}{2004}) 
  {ArXiv:astro-ph/0304268}.
  
\bibitem[10]{seo2003}
\bibinfo{author}{\bibfnamefont{H.-J.} \bibnamefont{{Seo}}} \bibnamefont{and}
  \bibinfo{author}{\bibfnamefont{D.~J.} \bibnamefont{{Eisenstein}}},
  \bibinfo{journal}{\apj} \textbf{\bibinfo{volume}{598}}, \bibinfo{pages}{720}
  (\bibinfo{year}{2003})  {ArXiv:astro-ph/0307460}.


\bibitem[11]{complex_step}
\bibinfo{author}{\bibfnamefont{J.~R. R.~A.} \bibnamefont{Martins}},
  \bibinfo{author}{\bibfnamefont{P.}~\bibnamefont{Sturdza}}, \bibnamefont{and}
  \bibinfo{author}{\bibfnamefont{J.~J.} \bibnamefont{Alonso}},
  \bibinfo{journal}{ACM Trans. Math. Softw.} \textbf{\bibinfo{volume}{29}},
  \bibinfo{pages}{245} (\bibinfo{year}{2003}).


\bibitem[12]{cp}
\bibinfo{author}{\bibfnamefont{M.}~\bibnamefont{{Chevallier}}}
  \bibnamefont{and}
  \bibinfo{author}{\bibfnamefont{D.}~\bibnamefont{{Polarski}}},
  \bibinfo{journal}{Int. J. Mod. Phys. D} \textbf{\bibinfo{volume}{10}},
  \bibinfo{pages}{213} (\bibinfo{year}{2001}).


\bibitem[13]{linder_w}
\bibinfo{author}{\bibfnamefont{E.~V.} \bibnamefont{{Linder}}},
  \bibinfo{journal}{Physical Review Letters} \textbf{\bibinfo{volume}{90}},
  \bibinfo{pages}{091301} (\bibinfo{year}{2003})  {ArXiv:astro-ph/0208512}.

\bibitem[14]{correct_detf}
\bibinfo{author}{\bibfnamefont{A.}~\bibnamefont{{Albrecht}}},
  \bibinfo{author}{\bibfnamefont{G.}~\bibnamefont{{Bernstein}}},
  \bibinfo{author}{\bibfnamefont{R.}~\bibnamefont{{Cahn}}},
  \bibnamefont{et~al.} (\bibinfo{year}{2006})  {ArXiv:astro-ph/0609591}.

\bibitem[15]{amendola}
\bibinfo{author}{\bibfnamefont{L.}~\bibnamefont{{Amendola}}},
  \bibinfo{author}{\bibfnamefont{C.}~\bibnamefont{{Quercellini}}},
  \bibnamefont{and}
  \bibinfo{author}{\bibfnamefont{E.}~\bibnamefont{{Giallongo}}},
  \bibinfo{journal}{\mnras} \textbf{\bibinfo{volume}{357}},
  \bibinfo{pages}{429} (\bibinfo{year}{2005})  {ArXiv:astro-ph/0404599}.
  
\bibitem[16]{amendola07}
\bibinfo{author}{\bibfnamefont{C.}~\bibnamefont{{Di Porto}}} \bibnamefont{and}
  \bibinfo{author}{\bibfnamefont{L.}~\bibnamefont{{Amendola}}} (\bibinfo{year}{2007}) {ArXiv:0707.2686}.
  
\bibitem[17]{loeb_growth}
\bibinfo{author}{\bibfnamefont{A.}~\bibnamefont{{Loeb}}} \bibnamefont{and}
  \bibinfo{author}{\bibfnamefont{S.}~\bibnamefont{{Wyithe}}}  (\bibinfo{year}{2008}) {ArXiv:0801.1677}.

\bibitem[18]{wang_growth}
\bibinfo{author}{\bibfnamefont{Y.}~\bibnamefont{{Wang}}}(\bibinfo{year}{2007})  {ArXiv:0710.3885}.

\bibitem[19]{linder_growth05}
\bibinfo{author}{\bibfnamefont{E.~V.} \bibnamefont{{Linder}}},
  \bibinfo{journal}{\prd} \textbf{\bibinfo{volume}{72}},
  \bibinfo{pages}{043529} (\bibinfo{year}{2005}{\natexlab{a}}) 
  {ArXiv:astro-ph/0507263}.

\bibitem[20]{linder_growth09}
\bibinfo{author}{\bibfnamefont{E.~V.} \bibnamefont{{Linder}}},
  \bibinfo{journal}{\prd} \textbf{\bibinfo{volume}{79}},
  \bibinfo{pages}{063519} (\bibinfo{year}{2009})  {ArXiv:0901.0918}.

\bibitem[21]{peebles1993}
\bibinfo{author}{\bibfnamefont{P.~J.~E.} \bibnamefont{{Peebles}}},
  \emph{\bibinfo{title}{{Principles of physical cosmology}}}, Princeton
  University Press (\bibinfo{year}{1993}).

\bibitem[22]{wang_steinhardt}
\bibinfo{author}{\bibfnamefont{L.}~\bibnamefont{{Wang}}} \bibnamefont{and}
  \bibinfo{author}{\bibfnamefont{P.~J.} \bibnamefont{{Steinhardt}}},
  \bibinfo{journal}{\apj} \textbf{\bibinfo{volume}{508}}, \bibinfo{pages}{483}
  (\bibinfo{year}{1998})  {ArXiv:astro-ph/9804015}.

\bibitem[23]{linder_jenkins}
\bibinfo{author}{\bibfnamefont{E.~V.} \bibnamefont{{Linder}}} \bibnamefont{and}
  \bibinfo{author}{\bibfnamefont{A.}~\bibnamefont{{Jenkins}}},
  \bibinfo{journal}{\mnras} \textbf{\bibinfo{volume}{346}},
  \bibinfo{pages}{573} (\bibinfo{year}{2003})  {ArXiv:astro-ph/0305286}.

\bibitem[24]{growth_eis}
\bibinfo{author}{\bibfnamefont{D.~J.} \bibnamefont{{Eisenstein}}} (\bibinfo{year}{1997}) 
  {ArXiv:astro-ph/9709054}.

\bibitem[25]{heath77}
\bibinfo{author}{\bibfnamefont{D.~J.} \bibnamefont{{Heath}}},
  \bibinfo{journal}{\mnras} \textbf{\bibinfo{volume}{179}},
  \bibinfo{pages}{351} (\bibinfo{year}{1977}).

\bibitem[26]{blake_etal_ff}
\bibinfo{author}{\bibfnamefont{C.}~\bibnamefont{{Blake}}},
  \bibinfo{author}{\bibfnamefont{D.}~\bibnamefont{{Parkinson}}},
  \bibinfo{author}{\bibfnamefont{B.}~\bibnamefont{{Bassett}}},
  \bibnamefont{et~al.}, \bibinfo{journal}{\mnras}
  \textbf{\bibinfo{volume}{365}}, \bibinfo{pages}{255} (\bibinfo{year}{2006}) 
  {ArXiv:astro-ph/0510239}.
  
  
\bibitem[27]{SE2007}
\bibinfo{author}{\bibfnamefont{H.-J.} \bibnamefont{{Seo}}} \bibnamefont{and}
  \bibinfo{author}{\bibfnamefont{D.~J.} \bibnamefont{{Eisenstein}}},
  \bibinfo{journal}{\apj} \textbf{\bibinfo{volume}{665}}, \bibinfo{pages}{14}
  (\bibinfo{year}{2007})  {ArXiv:astro-ph/0701079}.


\bibitem[28]{wangsh}
\bibinfo{author}{\bibfnamefont{Y.}~\bibnamefont{{Wang}}},
  \bibinfo{journal}{\apj} \textbf{\bibinfo{volume}{647}}, \bibinfo{pages}{1}
  (\bibinfo{year}{2006})  {ArXiv:astro-ph/0601163}.
  
  
\bibitem[29]{SE2007code}
\emph{\bibinfo{title}{http://cmb.as.arizona.edu/$\sim$eisenste/acousticpeak/bao\_fo%
recast.html}}.


\bibitem[30]{weinberg1972}
\bibinfo{author}{\bibfnamefont{S.}~\bibnamefont{{Weinberg}}},
  \emph{\bibinfo{title}{{Gravitation and Cosmology: Principles and Applications
  of the General Theory of Relativity}}}, Wiley-VCH (\bibinfo{year}{1972}).

\bibitem[31]{hogg}
\bibinfo{author}{\bibfnamefont{D.~W.} \bibnamefont{{Hogg}}} (\bibinfo{year}{1999}) 
  {ArXiv:astro-ph/9905116}.

\bibitem[32]{wmap5}
\bibinfo{author}{\bibfnamefont{E.}~\bibnamefont{{Komatsu}}},
  \bibinfo{author}{\bibfnamefont{J.}~\bibnamefont{{Dunkley}}},
  \bibinfo{author}{\bibfnamefont{M.~R.} \bibnamefont{{Nolta}}},
  \bibnamefont{et~al.}, \bibinfo{journal}{\apjs}
  \textbf{\bibinfo{volume}{180}}, \bibinfo{pages}{330} (\bibinfo{year}{2009}) 
  {ArXiv:0803.0547}.

\bibitem[33]{Carroll_1992}
\bibinfo{author}{\bibfnamefont{S.~M.} \bibnamefont{{Carroll}}},
  \bibinfo{author}{\bibfnamefont{W.~H.} \bibnamefont{{Press}}},
  \bibnamefont{and} \bibinfo{author}{\bibfnamefont{E.~L.}
  \bibnamefont{{Turner}}}, \bibinfo{journal}{\araa}
  \textbf{\bibinfo{volume}{30}}, \bibinfo{pages}{499} (\bibinfo{year}{1992}).

\bibitem[34]{lahav_suto}
\bibinfo{author}{\bibfnamefont{O.}~\bibnamefont{{Lahav}}} \bibnamefont{and}
  \bibinfo{author}{\bibfnamefont{Y.}~\bibnamefont{{Suto}}},
  \bibinfo{journal}{Living Reviews in Relativity} \textbf{\bibinfo{volume}{7}},
  \bibinfo{pages}{8} (\bibinfo{year}{2004})  {ArXiv:astro-ph/0310642}.

\bibitem[35]{peacock_dodds}
\bibinfo{author}{\bibfnamefont{J.~A.} \bibnamefont{{Peacock}}}
  \bibnamefont{and} \bibinfo{author}{\bibfnamefont{S.~J.}
  \bibnamefont{{Dodds}}}, \bibinfo{journal}{\mnras}
  \textbf{\bibinfo{volume}{267}}, \bibinfo{pages}{1020} (\bibinfo{year}{1994}) 
  {ArXiv:astro-ph/9311057}.

\end{thebibliography}

\end{document}